\documentclass[journal,twocolumn,10pt,twoside]{IEEEtran}
\newtheorem{proposition}{{Proposition}}
\newtheorem{definition}{{Definition}}
\newtheorem{theorem}{{Theorem}}
\newtheorem{lemma}{{Lemma}}

\newtheorem{corollary}{{Corollary}}
\newtheorem{remark}{{Remark}}

\usepackage{cite}
\usepackage{amsmath,amssymb,amsfonts}
\usepackage{algorithmic}
\usepackage{graphicx}
\usepackage{textcomp}
\usepackage{xcolor}

\normalsize

\ifCLASSINFOpdf
\else
\fi

\begin{document}
%
\title{Output-Constrained Lossy Source Coding With Application to Rate-Distortion-Perception Theory}
%
%
%

\author{Li~Xie, Liangyan~Li,        Jun~Chen, 
        and~Zhongshan~Zhang
}

\maketitle

\begin{abstract}
The distortion-rate function of output-constrained lossy source coding with limited common randomness is analyzed for the special case of squared error distortion measure. An explicit expression is obtained when both source and reconstruction distributions are Gaussian. This further leads to a partial characterization of 
the information-theoretic limit of quadratic Gaussian rate-distortion-perception coding with the perception measure given by 
Kullback-Leibler divergence or squared quadratic Wasserstein distance.
\end{abstract}


%
\IEEEpeerreviewmaketitle

\section{Introduction}

\IEEEPARstart{I}{t} has been long recognized in image compression that the perceptual quality of a compressed image is not completely aligned its distortion with respect to the original version. Indeed, different from the full-reference nature of distortion measure, perception measure is more concerned with the difference in the statistical properties than the actual pixel values. By viewing each image as a random sample from a certain distribution that encodes its statistical properties, perceptual quality assessment can be performed by comparing pre- and post-compression image distributions. 

Equipped with properly defined distortion and perception measures,  Blau and Michaeli \cite{BM18} demonstrated quantatively
the tension between reconstruction distortion and  perceptual quality through an empirical investigation of  GAN-based image restoration algorithms. In \cite{BM19}, they further initiated a theoretical study of the three-way tradeoff among compression rate, reconstruction distortion, perceptual quality; in particular, a rate-distortion-perception function was defined by generalizing Shannon's rate-distortion function and was conjectured to characterize the information-theoretic limit of the aforementioned three-way tradeoff (see also \cite{Matsumoto18,Matsumoto19} for some related work). Theis and Wagner \cite{TW21} proved this conjecture by allowing the encoder and decoder to have access to unlimited common randomness. Later, Chen et al. \cite{CYWSGT22} showed that the rate-distortion-perception function introduced by Blau and Michaeli can still be achieved even when no common randomness is available. However, in contrast to \cite{TW21}, the coding theorem in \cite{CYWSGT22} does not ensure that the reconstructed symbols are independent and identically distributed (i.i.d.); for this reason, it is impossible to enforce the sequence-level distributional consistency between the source and reconstruction using the marginal-distribution-based perception constraint as adopted in the original problem formulation
\cite{BM19}.
On the other hand, as shown by Wagner \cite{Wagner22}, generally a price has to be paid in terms of the rate-distortion tradeoff to maintain the perfect sequence-level distributional consistency between the source and reconstruction with no or limited common randomness.

In this work, we aim to further study the impact of common randomness on the fundamental limit of rate-distortion-perception coding
by exploring its connection with output-constrained lossy source coding \cite{LKK10,LKK11,KZLK13,SLY15J1,SLY15J2}.
Output-constrained lossy source coding differs from conventional lossy source coding in the sense that the reconstructed symbols are required to be i.i.d. with a prescribed marginal distribution. This formulation is well suited to our purpose as it enables us to gain an effective control of the sequence-level distributional difference between the source and reconstruction via the marginal-distribution-based perception constraint. Moreover, the role of common randomness in output-constrained lossy source coding  is largely understood \cite{SLY15J2}.  
On the other hand, to explicitly characterize the dependency of the rate-distortion-perception tradeoff on the available amount of common randomness, one has to identify the optimal marginal distribution for the reconstruction sequence given the perception constraint and evaluate the corresponding information-theoretic limit of output-constrained lossy source coding, which is a non-trivial task in general. We make some progress in this regard by developing a systematic approach for the quadratic Gaussian case with two commonly used perception measures.





The rest of this paper is organized as follows. In Section \ref{sec:definition}, we link the information-theoretic limit of rate-distortion-perception coding to that of output-constrained lossy source coding based on their operational definitions. Section 
\ref{sec:general} presents a new characterization of the distortion-rate function of output-constrained lossy source coding under squared error distortion measure, through which some bounds are established. We show in Section \ref{sec:Gaussian} that these bounds coincide when both source and reconstruction distributions are Gaussian, and further leverage them to partially characterize the information-theoretic limit of quadratic Gaussian rate-distortion-perception coding with the perception measure given by Kullback-Leibler divergence or squared quadratic Wasserstein distance.
Section \ref{sec:conclusion} contains some concluding remarks.

We adopt the conventional notation for information measures: $H(\cdot)$ for entropy, $h(\cdot)$ for differential entropy, and $I(\cdot;\cdot)$ for mutual information. The  distribution, mean, and variance of random variable $X$ are denoted by $p_X$, $\mu_X$, and $\sigma^2_X$, respectively. The cardinality of set $\mathcal{S}$ is written as $|\mathcal{S}|$. 
For any real numbers $a$ and $b$, we use $(a)_+$, $a\vee b$, and $a\wedge b$ to represent  $\max\{a,0\}$, $\max\{a,b\}$, and $\min\{a,b\}$, respectively. Throughout this paper, the base of the logarithm
function is assumed to be $e$.



\section{Problem Definition}\label{sec:definition}

\begin{figure}[htbp]
	\centerline{\includegraphics[width=9cm]{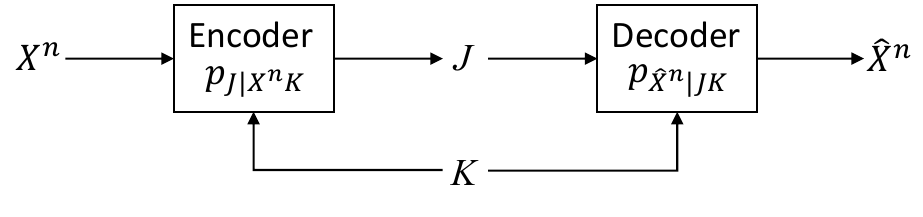}} \caption{System diagram.}
	\label{fig:system} 
\end{figure}

Let the source $\{X_t\}_{t=1}^{\infty}$ be a stationary and memoryless process with marginal distribution $p_X$ over alphabet $\mathcal{X}$. Each length-$n$ output-constrained lossy source coding system (see Fig. \ref{fig:system}) consists of a stochastic encoder $f^{(n)}:\mathcal{X}^n\times\mathcal{K}\rightarrow\mathcal{J}$, a stochastic decoder $g^{(n)}:\mathcal{J}\times\mathcal{K}\rightarrow\mathcal{X}^n$, and a shared random seed $K$, which is uniformly distributed over $\mathcal{K}$ and independent of the source.
The stochastic encoder $f^{(n)}$ maps  source sequence $X^n$ and random seed $K$ to a codeword $J$ in  $\mathcal{J}$ according to some conditional distribution $p_{J|X^nK}$. The stochastic decoder $g^{(n)}$ generates a reconstruction sequence $\hat{X}^n$ based on $J$ and $K$ according to some conditional distribution $p_{\hat{X}^n|JK}$. It is required that $\hat{X}^n$ is a sequence of i.i.d. random variables with a prescribed marginal distribution $p_{\hat{X}}$.
Note that the whole system is fully specified by the joint distribution $p_{X^n}p_Kp_{J|X^nK}p_{\hat{X}^n|JK}$.
Let $\Delta:\mathcal{X}\times\mathcal{X}\rightarrow[0,\infty)$ be a distortion measure.
The end-to-end distortion of the above coding system is quantified by $\frac{1}{n}\sum_{t=1}^n\mathbb{E}[\Delta(X_t,\hat{X}_t)]$.

\begin{definition}
	Distortion level $D$ is said to be achievable  with respect to reconstruction distribution $p_{\hat{X}}$ subject to rate constraints $R$ and $R_c$ 	
	if there exist encoder $f^{(n)}$ and decoder $g^{(n)}$ such that 	
	\begin{align*}
		&\frac{1}{n}\log|\mathcal{J}|\leq R,\\
		&\frac{1}{n}\log|\mathcal{K}|\leq R_c,\\
		&\frac{1}{n}\sum\limits_{t=1}^n\mathbb{E}[\Delta(X_t,\hat{X}_t)]\leq D,
	\end{align*}
and $\hat{X}^n$ is a sequence of i.i.d. random variables with marginal distribution $p_{\hat{X}}$. The infimum of such achievable $D$ is denoted by  $D(R,R_c|p_X,p_{\hat{X}})$.
	\end{definition}

Rate-distortion-perception coding is similar to output-constrained lossy source coding except that the reconstruction distribution $p_{\hat{X}}$, instead of being predetermined, is only required to be close to the source distribution $p_X$ under certain perception measure $\phi$. Specifically, $\phi:\mathcal{P}\times\mathcal{P}\rightarrow[0,\infty]$
is a divergence with $\phi(p_X,p_{\hat{X}})=0$ if and only if $p_X=p_{\hat{X}}$, where $\mathcal{P}$ denotes the set of probability distributions;  the perceptual quality of the coding system is quantified by $\frac{1}{n}\sum_{t=1}^n\phi(p_{X_t},p_{\hat{X}_t})$.

\begin{figure*}[htbp]
	\centerline{\includegraphics[width=18cm]{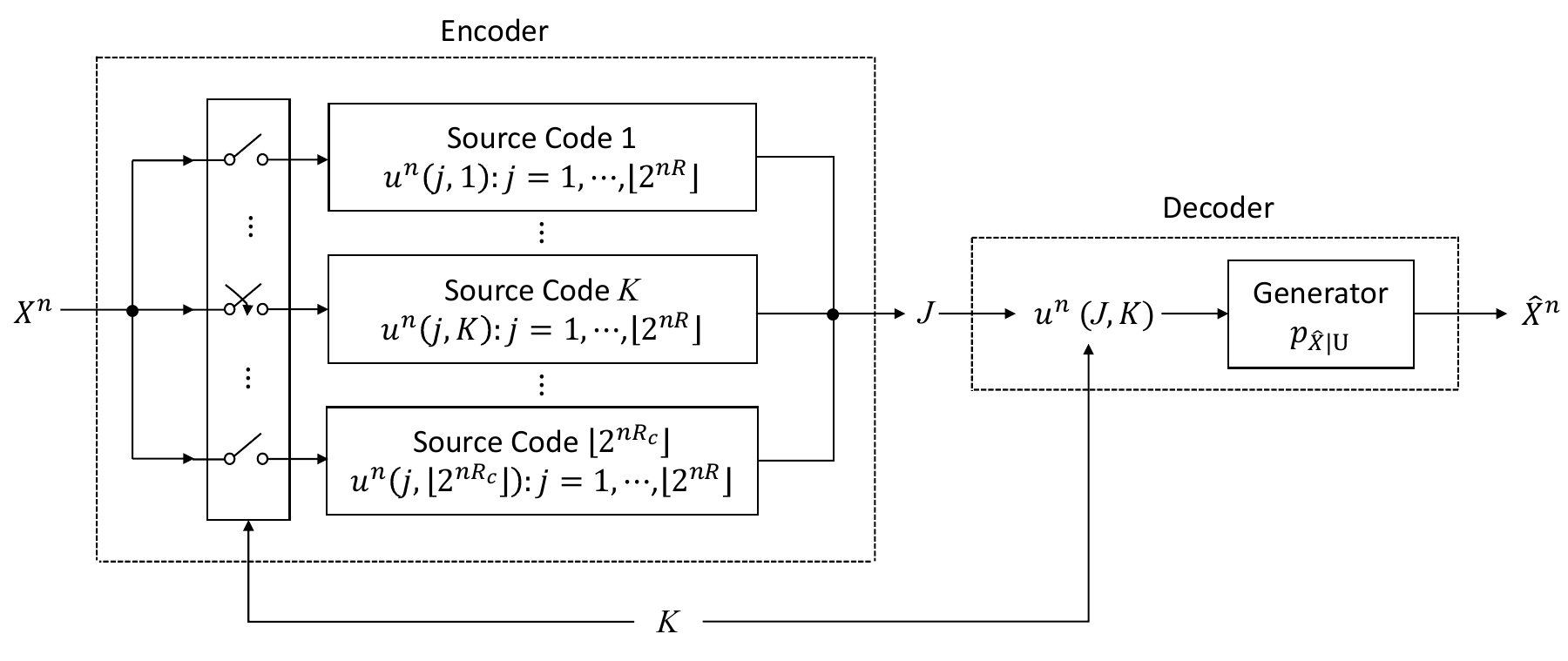}} \caption{A generic coding scheme.}
	\label{fig:slides-1} 
\end{figure*}

\begin{definition}
		Distortion level $D$ is said to be achievable   subject to rate constraints $R$ and $R_c$ as well as perception constraint $P$	
	if there exist encoder $f^{(n)}$ and decoder  $g^{(n)}$ such that 	
	\begin{align}
		&\frac{1}{n}\log|\mathcal{J}|\leq R,\nonumber\\
		&\frac{1}{n}\log|\mathcal{K}|\leq R_c,\nonumber\\
		&\frac{1}{n}\sum\limits_{t=1}^n\mathbb{E}[\Delta(X_t,\hat{X}_t)]\leq D,\nonumber\\
		&\frac{1}{n}\sum\limits_{t=1}^n\phi(p_{X_t},p_{\hat{X}_t})\leq P,\label{eq:marginal_perception}
	\end{align}
	and $\hat{X}^n$ is a sequence of i.i.d. random variables. The infimum of such achievable $D$ is denoted by  $D(R,R_c,P|p_X)$.
\end{definition}

Since the source variables are identically distributed and so are the reconstruction variables, $\phi(p_{X_t},p_{\hat{X}_t})$ actually does not depend on $t$. It is thus clear that rate-distortion-perception coding is equivalent to output-constrained source coding with the reconstruction distribution restricted to $\{p_{\hat{X}}:\phi(p_X,p_{\hat{X}})\leq P\}$. As a consequence, we have
\begin{align}
	D(R,R_c,P|p_X)&=\inf\limits_{p_{\hat{X}}:\phi(p_X,p_{\hat{X}})\leq P}D(R,R_c|p_X,p_{\hat{X}}).\label{eq:infpXhat}
\end{align}

There are many possible choices of perception measure. Of particular interest to us is the case $\phi(p_X,p_{\hat{X}})=\phi_{KL}(p_{\hat{X}}\|p_X)$, where
\begin{align*}
	\phi(p_{\hat{X}}\|p_X):=\mathbb{E}\left[\log\frac{p_{\hat{X}}(\hat{X})}{p_X(\hat{X})}\right]
\end{align*}
is the Kullback-Leibler divergence.
We will frequently use the following extremal property of Gaussian distribution with respect to $\phi_{KL}$, which follows from the identity
\begin{align*}
	&\phi_{KL}(p_{\hat{X}}\|\mathcal{N}(\mu_X,\sigma^2_X))\\&=-h(\hat{X})+\frac{1}{2}\log(2\pi\sigma^2_{X})+\frac{(\mu_X-\mu_{\hat{X}})^2+\sigma^2_{\hat{X}}}{2\sigma^2_X}
\end{align*}
and the fact that
$h(\hat{X})\leq\frac{1}{2}\log(2\pi e\sigma^2_{\hat{X}})$  with equality if and only if $\hat{X}\sim\mathcal{N}(\mu_{\hat{X}},\sigma^2_{\hat{X}})$\cite[Theorem 9.6.5]{CT91}.

\begin{proposition}\label{prop:KL}
	For $p_{\hat{X}}$ with $\mathbb{E}[\hat{X}^2]<\infty$, we have
	\begin{align*}
		&\phi_{KL}(p_{\hat{X}}\|\mathcal{N}(\mu_X,\sigma^2_X))\\
		&\geq\phi_{KL}(\mathcal{N}(\mu_{\hat{X}},\sigma^2_{\hat{X}})\|\mathcal{N}(\mu_X,\sigma^2_X))\\
		&=\log\frac{\sigma_X}{\sigma_{\hat{X}}}+\frac{(\mu_X-\mu_{\hat{X}})^2+\sigma^2_{\hat{X}}-\sigma^2_X}{2\sigma^2_X}.
	\end{align*}
\end{proposition}

Another perception measure of interest to us is the squared quadratic Wasserstein distance
\begin{align}
	W^2_2(p_X,p_{\hat{X}}):=\inf\limits_{p_{X\hat{X}}\in\Pi(p_X,p_{\hat{X}})}\mathbb{E}[(X-\hat{X})^2]\label{eq:inf},
\end{align}
where $\Pi(p_X,p_{\hat{X}})$ denotes the set of all possible couplings of $p_X$ and $p_{\hat{X}}$. The following result \cite[Equation (6)]{DL82} 
\cite[Proposition 7]{GS84} indicates that the extremal property of Gaussian distribution also manifests under $W^2_2$.

\begin{proposition}\label{prop:W2}
	For $p_{X}$ and $p_{\hat{X}}$ with $\mathbb{E}[X^2]<\infty$ and $\mathbb{E}[\hat{X}^2]<\infty$, we have
\begin{align*}
	&W^2_2(p_X,p_{\hat{X}})\\
	&\geq W^2_2(\mathcal{N}(\mu_X,\sigma^2_X),\mathcal{N}(\mu_{\hat{X}},\sigma^2_{\hat{X}}))\\
	&=(\mu_X-\mu_{\hat{X}})^2+(\sigma_X-\sigma_{\hat{X}}).
\end{align*}
\end{proposition}

\section{General Case}\label{sec:general}

A tuple $(p_X, p_{\hat{X}}, \Delta)$ of source distribution,
reconstruction distribution, and distortion measure is
said to be uniformly integrable if for every $\epsilon>0$, there
exists $\delta>0$ such that $\sup_{p_{X\hat{X}},\mathcal{E}}\mathbb{E}[\Delta(X,\hat{X})1_{\mathcal{E}}(X,\hat{X})]\leq\epsilon$, where the supremum is over all $p_{X\mathcal{X}}\in\Pi(p_X,p_{\hat{X}})$ and
all measurable events $\mathcal{E}$ with $\mathbb{P}\{(X,\hat{X})\in\mathcal{E}\}\leq\delta$.
Moreover, let $\Omega(p_X,p_{\hat{X}})$ denote the set of joint distributions $p_{XU\hat{X}}$ compatible with the given marginals $p_X$ and $p_{\hat{X}}$ such that $X\leftrightarrow U\leftrightarrow\hat{X}$ form a Markov chain.

\begin{figure*}[htbp]
	\centerline{\includegraphics[width=18cm]{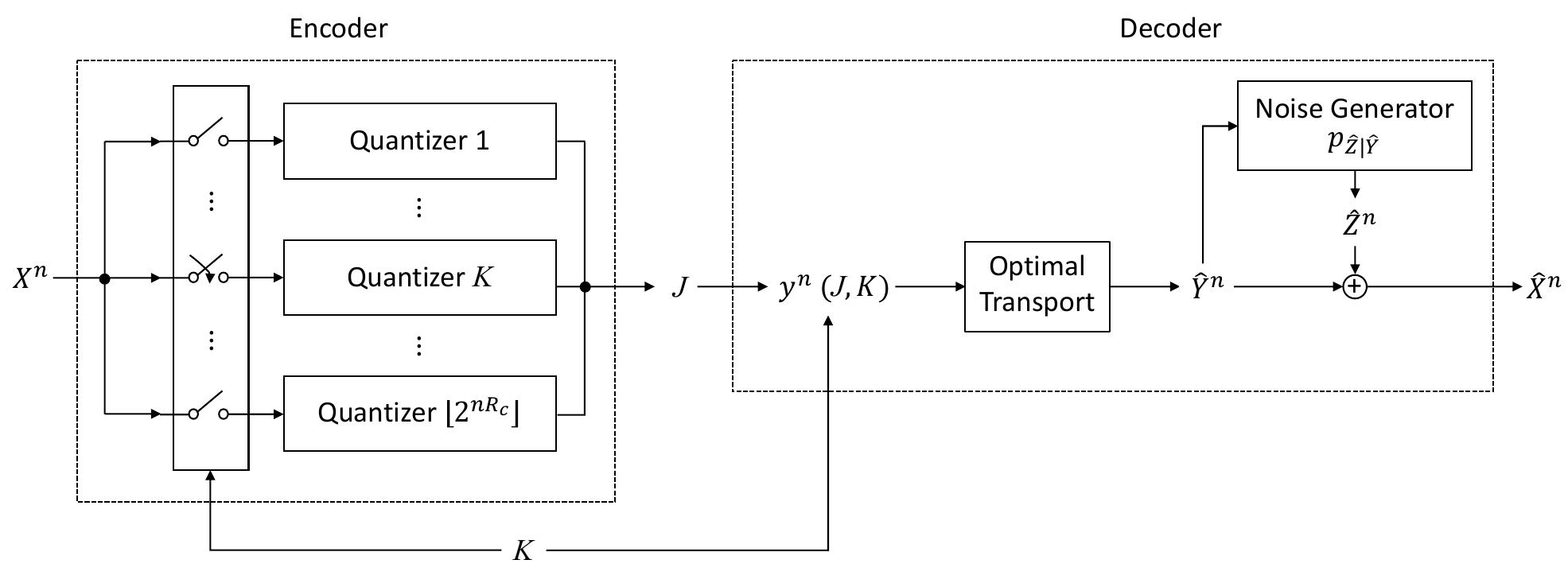}} \caption{A specialized coding scheme for squared error distortion measure.}
	\label{fig:slides-2} 
\end{figure*}

\begin{theorem}\label{thm:outputconstrained}
	If $(p_X,p_{\hat{X}},\Delta)$ is uniformly integrable, then
	\begin{align}
		D(R,R_c|p_X,p_{\hat{X}})&=\inf\limits_{p_{XU\hat{X}}\in\Omega(p_X,p_{\hat{X}})}\mathbb{E}[\Delta(X,\hat{X})]\label{eq:expression1}\\
		&\mbox{subject to }\quad I(X;U)\leq R,\label{eq:constrainta}\\
		&\hspace{0.74in} I(\hat{X};U)\leq R+R_c.\label{eq:constraintb}
	\end{align}
	Moreover, for the case   $\mathbb{E}[X^2]<\infty$, $\mathbb{E}[\hat{X}^2]<\infty$, and $\Delta(x,\hat{x})=(x-\hat{x})^2$, we have
	\begin{align}
		&D(R,R_c|p_X,p_{\hat{X}})\nonumber\\
		&=\inf\limits_{p_{Y|X},p_{\hat{Y}|\hat{X}}}\mathbb{E}[(X-Y)^2]+\mathbb{E}[(\hat{X}-\hat{Y})^2]+W^2_2(p_{Y},p_{\hat{Y}})\label{eq:expression2}\\
		&\mbox{subject to }\quad\mathbb{E}[X|Y]=Y\mbox{ a.s.},\label{eq:constraint1}\\
		&\hspace{0.74in} \mathbb{E}[\hat{X}|\hat{Y}]=\hat{Y}\mbox{ a.s.},\label{eq:constraint2}\\
		&\hspace{0.74in} I(X;Y)\leq R,\label{eq:constraint3}\\
		&\hspace{0.74in} I(Y;\hat{Y})\leq R+R_c.\label{eq:constraint4}
	\end{align}	
\end{theorem}
\begin{IEEEproof}
	See Appendix \ref{app:outputconstrained}.
\end{IEEEproof}
\begin{remark}
The coding scheme associated with the single-letter charaterization in (\ref{eq:expression1})--(\ref{eq:constraintb}) can be roughly described as follows (see Fig. \ref{fig:slides-1}). First construct  $\lfloor 2^{nR_c}\rfloor\approx 2^{n(I(\hat{X};U)-I(X;U))}$ source codes, each with $\lfloor2^{nR}\rfloor\approx 2^{nI(X;U)}$ codewords. Given source sequence $X^n$, the encoder maps it to a codeword $u^n(J,K)$ in the source code specified by shared random seed $K$, and sends codeword index $J$ to the decoder. The decoder will view the $\lfloor 2^{nR_c}\rfloor$ source codes collectively as a source code, which consists of approximately $2^{nI(\hat{X};U)}$ codewords, and view $u^n(J,K)$ as the encoded version of reconstruction sequence $\hat{X}^n$ based on this code. Therefore, to generate $\hat{X}^n$ from $u^n(J,K)$, it needs to invert the lossy source encoding operation, which can be essentially realized by passing $u^n(J,K)$ through memoryless channel $p_{\hat{X}|U}$ \cite{ASP23,SCKY24}. Note that the encoder basically implements conventional lossy source encoding, which is a deterministic operation, whereas the decoder implements the stochastic inverse of lossy source encoding, which is in general not deterministic. 

For squared error distortion measure, in view of the single-letter characterization in (\ref{eq:expression2})--(\ref{eq:constraint4}), the above coding scheme can be specialized as follows (see Fig. \ref{fig:slides-2}). Here each source code becomes a quantizer. 
The encoder  quantizes source sequence $X^n$ using the quantizer specified by shared random seed $K$.
The decoder converts quantizatizer output $y^n(J,K)$  to another sequence $\hat{Y}^n$ in a symbol-wise manner via the optimal transport plan that achieves $W^2_2(p_Y,p_{\hat{Y}})$. It then adds noise sequence $\hat{Z}^n$ to $\hat{Y}^n$ to produce reconstruction sequence $\hat{X}^n$, where $\hat{Z}^n$ is generated based on $\hat{Y}^n$ through memoryless channel $p_{\hat{Z}|\hat{Y}}:=p_{\hat{X}-\hat{Y}|\hat{Y}}$. Some related results in the one-shot setting can be found in \cite{YWYML21,TA21,LZCK22,LZCK22J}.
\end{remark}




For $p_X$ and $p_{\hat{X}}$ with $\mathbb{E}[X^2]<\infty$ and $\mathbb{E}[\hat{X}^2]<\infty$, define
\begin{align*}
	&D(R|p_X):=\min\limits_{p_{Y|X}:I(X;Y)\leq R}\mathbb{E}[(X-Y)^2],\\
	&D(R+R_c|p_{\hat{X}}):=\min\limits_{p_{\hat{Y}|\hat{X}}:I(\hat{X};\hat{Y})\leq R+R_c}\mathbb{E}[(\hat{X}-\hat{Y})^2],
\end{align*}
which are respectively the quadratic distortion-rate functions of $p_X$ and $p_{\hat{X}}$ evaluated at $R$ and $R+R_c$. Moreover, let
$Y^*$ and $\hat{Y}^*$ denote the outputs of the corresponding optimal test channels, and let $Y^G\sim\mathcal{N}(\mu_{Y^*},\sigma^2_{Y^*})$ and $\hat{Y}^G\sim\mathcal{N}(\mu_{\hat{Y}^*},\sigma^2_{\hat{Y}^*})$ be their Gaussian counterparts.

\begin{corollary}\label{cor:generallowerbound}
	For the case $\mathbb{E}[X^2]<\infty$, $\mathbb{E}[\hat{X}^2]<\infty$, and $\Delta(x,\hat{x})=(x-\hat{x})^2$, we have
	\begin{align*}
		\underline{D}(R,R_c|p_X,p_{\hat{X}})\leq D(R,R_c|p_X,p_{\hat{X}})\leq\overline{D}(R,R_c|p_X,p_{\hat{X}})
	\end{align*}
	where
	\begin{align*}
		&\overline{D}(R,R_c|p_X,p_{\hat{X}})\\
		&:=D(R|p_X)+D(R+R_c|p_{\hat{X}})+W^2_2(p_{Y^*},p_{\hat{Y}^*}),\\
&\underline{D}(R,R_c|p_X,p_{\hat{X}})\\
&:=D(R|p_X)+D(R+R_c|p_{\hat{X}})+W^2_2(p_{Y^G},p_{\hat{Y}^G}).
	\end{align*}	
\end{corollary}
\begin{IEEEproof}
	See Appendix \ref{app:generallowerbound}.	
\end{IEEEproof}
\begin{remark}
	The proof of Corollary \ref{cor:generallowerbound} indicates that $\underline{D}(R,R_c|p_X,p_{\hat{X}})$ can be written equivalently as
	\begin{align}
		&\underline{D}(R,R_c|p_X,p_{\hat{X}})
		\nonumber\\
		&=(\mu_X-\mu_{\hat{X}})^2+\sigma^2_X+\sigma^2_{\hat{X}}\nonumber\\
		&\quad-2\sqrt{(\sigma_X^2-D(R|p_X))(\sigma^2_{\hat{X}}-D(R+R_c|p_{\hat{X}}))}. \label{eq:remark}
	\end{align}
		One can obtain a more explicit lower bound on $D(R,R_c|p_X,p_{\hat{X}})$ by replacing $D(R|p_X)$ and $D(R+R_c|p_{\hat{X}})$ in (\ref{eq:remark}) with their respective Shannon lower bounds \cite[Equation (13.159)]{CT91}
	\begin{align}
		&D(R|p_X)\geq\frac{1}{2\pi e}e^{-2(R-h(X))},\nonumber\\
		&D(R+R_c|p_{\hat{X}})\geq\frac{1}{2\pi e}e^{-2(R+R_c-h(\hat{X}))}.\label{eq:Shannonlowerbound}
	\end{align}
	The condition $\mathbb{E}[X^2]<\infty$ and $\mathbb{E}[\hat{X}^2]<\infty$ ensures that $h(X)$ and $h(\hat{X})$ are well defined for $X$ and $\hat{X}$ with propbability densities \cite[Proposition 1]{Rioul11}. We set $h(X)=-\infty$  if $X$ does not have a probability density, and similarly for $h(\hat{X})$.
\end{remark}

\section{Gaussian Case}\label{sec:Gaussian}

\begin{figure}[htbp]
	\centerline{\includegraphics[width=9cm]{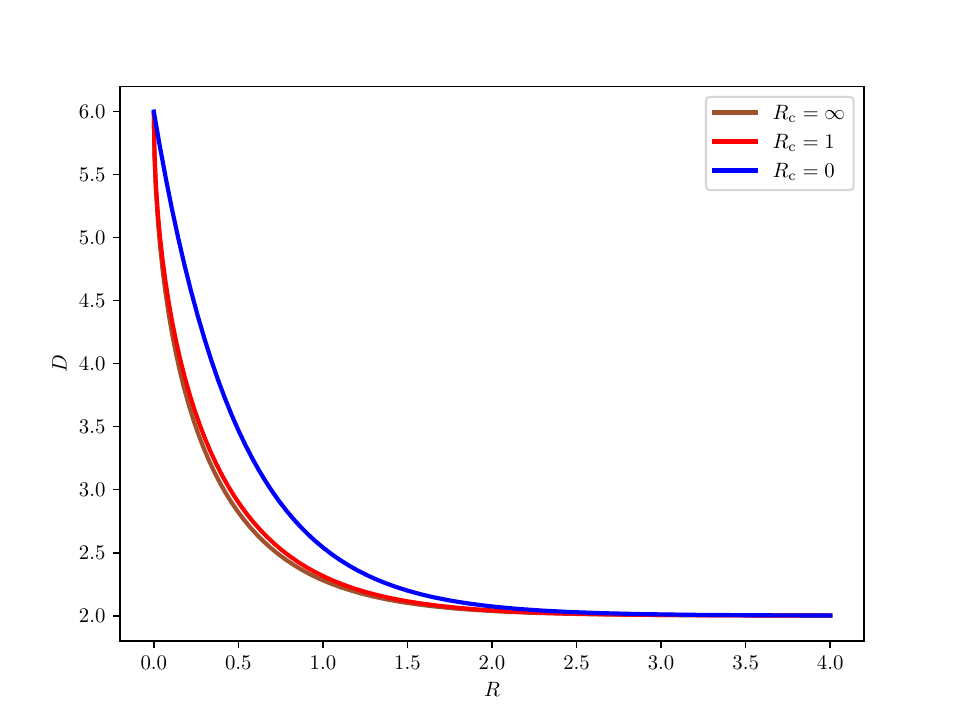}} \caption{Plots of $D(R,R_c|\mathcal{N}(0,1),\mathcal{N}(1,4))$ with $R_c=0, 1, \infty$.}
	\label{fig:Theorem2} 
\end{figure}

\begin{theorem}\label{thm:Gaussianoutputconstrained}
	For the case  $\Delta(x,\hat{x})=(x-\hat{x})^2$, we have
	\begin{align*}
		&D(R,R_c|\mathcal{N}(\mu_X,\sigma^2_X),\mathcal{N}(\mu_{\hat{X}},\sigma^2_{\hat{X}}))\\
		&=(\mu_X-\mu_{\hat{X}})^2+\sigma^2_X+\sigma^2_{\hat{X}}-2\sigma_X\sigma_{\hat{X}}\xi(R,R_c),
		\end{align*}
	where 
	\begin{align*}
		\xi(R,R_c):=\sqrt{(1-e^{-2R})(1-e^{-2(R+R_c)})}.
	\end{align*}
\end{theorem}
\begin{IEEEproof}
See Appendix \ref{app:Gaussianoutputconstrained}	
\end{IEEEproof}
\begin{remark}
Note that
\begin{align*}
	&D(0,R_c|\mathcal{N}(\mu_X,\sigma^2_X),\mathcal{N}(\mu_{\hat{X}},\sigma^2_{\hat{X}}))\\
	&=(\mu_X-\mu_{\hat{X}})^2+\sigma^2_X+\sigma^2_{\hat{X}},
\end{align*}
which is the mean squared error between $X$ and $\hat{X}$ when they are independent. On the other hand,
\begin{align*}
	&D(\infty,R_c|\mathcal{N}(\mu_X,\sigma^2_X),\mathcal{N}(\mu_{\hat{X}},\sigma^2_{\hat{X}}))\\
	&=(\mu_X-\mu_{\hat{X}})^2+(\sigma_X-\sigma_{\hat{X}})^2,
\end{align*}
which coincides with $W^2_2(\mathcal{N}(\mu_X,\sigma^2_X),\mathcal{N}(\mu_{\hat{X}},\sigma^2_{\hat{X}}))$. Moreover, it can be seen that $D(R,R_c|\mathcal{N}(\mu_X,\sigma^2_X),\mathcal{N}(\mu_{\hat{X}},\sigma^2_{\hat{X}}))$ is strictly decreasing in $R_c$ for a fixed $R$. Therefore, the availability of common randomness can affect the distortion-rate tradeoff in output-constrained lossy source coding. See Fig. \ref{fig:Theorem2} for the plots of $D(R,R_c|\mathcal{N}(0,1),\mathcal{N}(1,4))$ with $R_c=0, 1, \infty$, which indicate that a small amount of common randomness is almost as effective as unlimited common randomness.
\end{remark}

\begin{figure}[htbp]
	\centerline{\includegraphics[width=9cm]{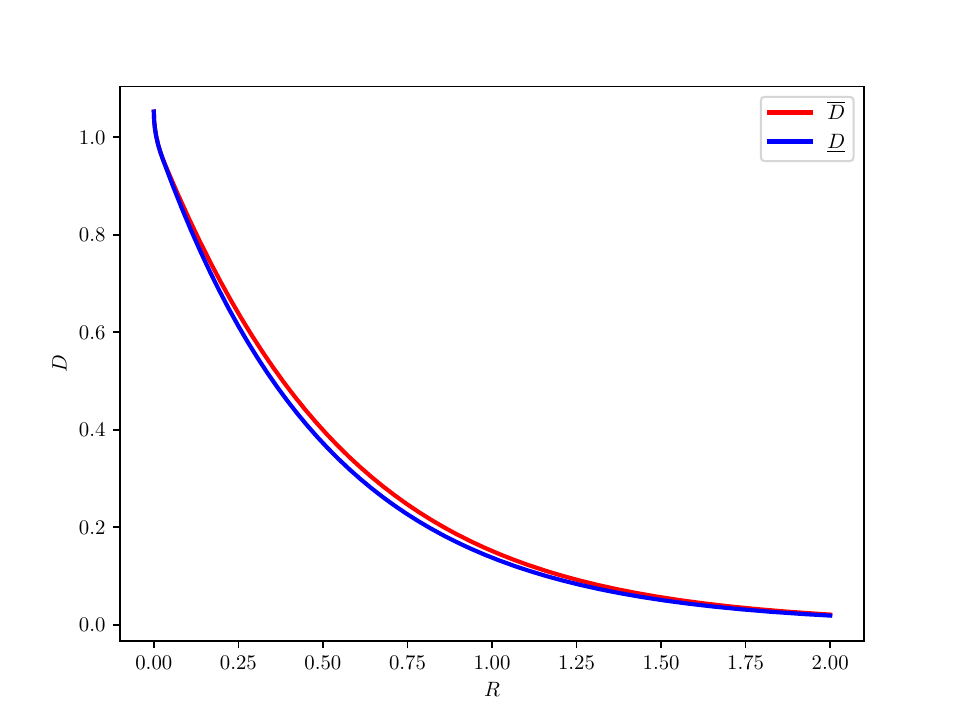}} \caption{Plots of $\overline{D}(R,1,1|\mathcal{N}(0,1))$ and $\underline{D}(R,1,1|\mathcal{N}(0,1))$.}
	\label{fig:Theorem3RD} 
\end{figure}

\begin{figure}[htbp]
	\centerline{\includegraphics[width=9cm]{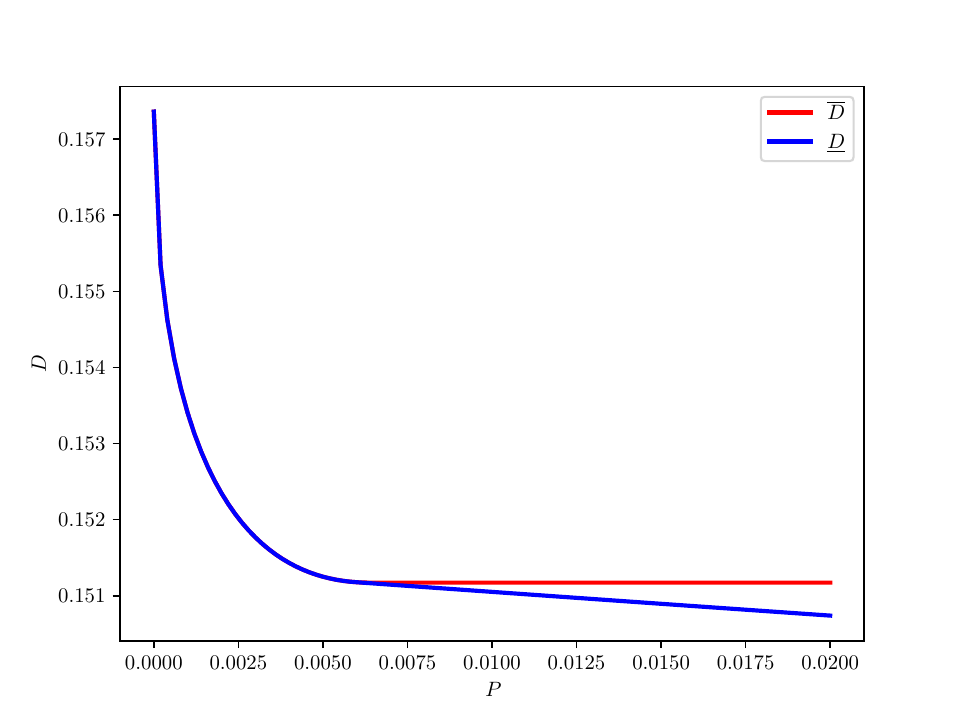}} \caption{Plots of $\overline{D}(1,1,P|\mathcal{N}(0,1))$ and $\underline{D}(1,1,P|\mathcal{N}(0,1))$.}
	\label{fig:Theorem3PD} 
\end{figure}

\begin{theorem}\label{thm:GaussianRDP}
	For the case  $\Delta(x,\hat{x})=(x-\hat{x})^2$ and $\phi(p_X,p_{\hat{X}})=\phi_{KL}(p_{\hat{X}}\|p_X)$, we have
	\begin{align}
		&D(R,R_c,P|\mathcal{N}(\mu_X,\sigma^2_X))\leq\overline{D}(R,R_c,P|\mathcal{N}(\mu_X,\sigma^2_X)),\label{eq:Gaussianupper}\\
&D(R,R_c,P|\mathcal{N}(\mu_X,\sigma^2_X))\geq\underline{D}(R,R_c,P|\mathcal{N}(\mu_X,\sigma^2_X)),\label{eq:Gaussianlower}
\end{align}
where
\begin{align*}
	&\overline{D}(R,R_c,P|\mathcal{N}(\mu_X,\sigma^2_X))\\
	&:=\begin{cases}
		\sigma^2_X-\sigma^2_X\xi^2(R,R_c)&\hspace{-0.7in}\mbox{if }\sigma(P)\leq\sigma_X\xi(R,R_c),\\
		\sigma^2_X+\sigma^2(P)-2\sigma_X\sigma(P)\xi(R,R_c)\\&\hspace{-0.7in}\mbox{if }\sigma(P)>\sigma_X\xi(R,R_c),
	\end{cases}\\
	&\underline{D}(R,R_c,P|\mathcal{N}(\mu_X,\sigma^2_X)):=\min\limits_{\sigma_{\hat{X}}\in[\sigma(P),\sigma_X]}\sigma^2_X+\sigma^2_{\hat{X}}\\
	&\hspace{0.5in}-2\sigma_X\sigma_{\hat{X}}\sqrt{(1-e^{-2R})(1-e^{-2(R+R_c+P-\psi(\sigma_{\hat{X}}))})}
\end{align*}
with 
\begin{align}
	\psi(\sigma_{\hat{X}}):=\log\frac{\sigma_X}{\sigma_{\hat{X}}}+\frac{\sigma^2_{\hat{X}}-\sigma^2_X}{2\sigma^2_X}\label{eq:defpsi}
\end{align}
and
$\sigma(P)$ being the unique number\footnote{  $\psi(\sigma)$ is a strictly descreasing function of $\sigma$ for $\sigma\in(0,\sigma^2_X]$ with $\psi(\sigma)\rightarrow\infty$ as $\sigma\rightarrow 0$ and $\psi(\sigma_X)=0$. So $\sigma(P)$ is uniquely defined for $P\in[0,\infty)$. We set $\sigma(P):=0$ when $P=\infty$.}  $\sigma\in[0,\sigma_X]$ satisfying $\psi(\sigma)=P$.
\end{theorem}
\begin{IEEEproof}
	See Appendix \ref{app:GaussianRDP}.
\end{IEEEproof}
\begin{remark}
	As shown by the plots of $\overline{D}(R,1,1|\mathcal{N}(0,1))$ and $\underline{D}(R,1,1|\mathcal{N}(0,1))$ in Fig. \ref{fig:Theorem3RD}, the two bounds are quite close to each other. In fact, they  coincide when $R$ is below a certain threshold. It can also be seen from Fig. \ref{fig:Theorem3PD} that $\overline{D}(1,1,P|\mathcal{N}(0,1))$ and $\underline{D}(1,1,P|\mathcal{N}(0,1))$ coincide when $P$ is below a certain threshold. 	
	These two phenomena turn out to be related as indicated by the following result.
	\end{remark}




\begin{figure}[htbp]
	\centerline{\includegraphics[width=9cm]{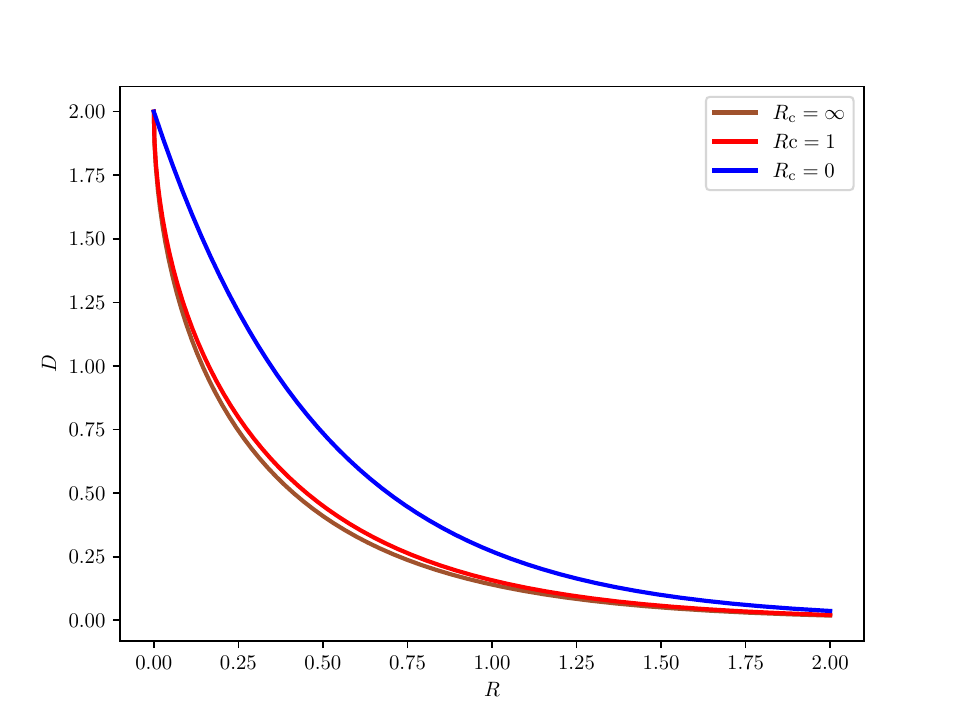}} \caption{Plots of $D(R,R_c,0|\mathcal{N}(0,1))$ with $R_c=0, 1, \infty$.}
	\label{fig:Corollary3Case3DR_rc1} 
\end{figure}

\begin{figure}[htbp]
	\centerline{\includegraphics[width=9cm]{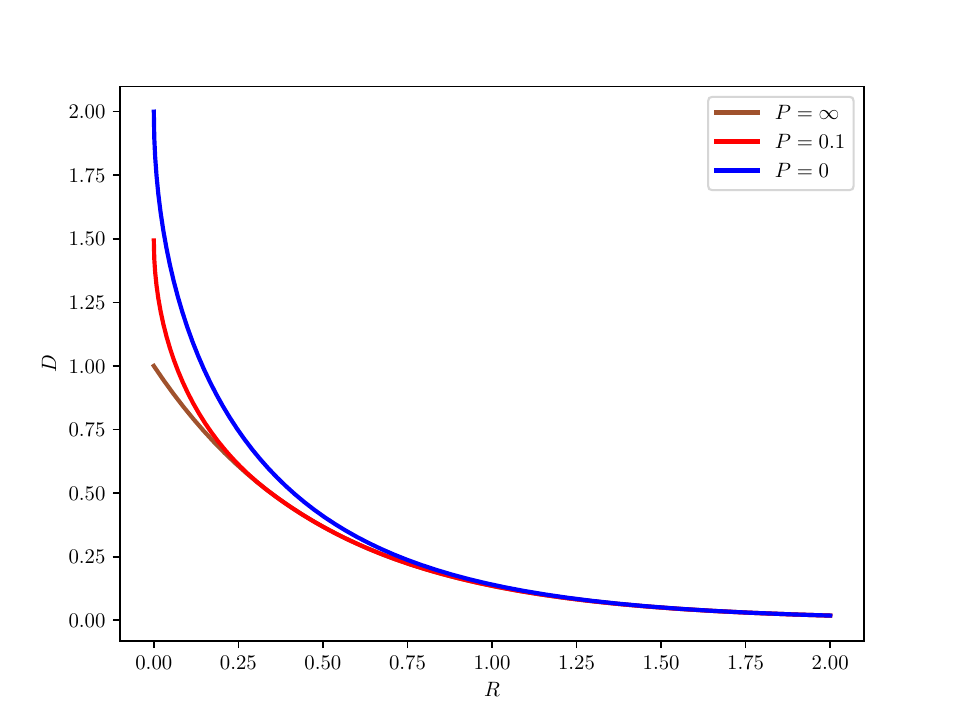}} \caption{Plots of $D(R,\infty,P|\mathcal{N}(0,1))$ with $P=0, 0.1, \infty$.}
	\label{fig:Corollary2case1DR} 
\end{figure}

\begin{figure}[htbp]
	\centerline{\includegraphics[width=9cm]{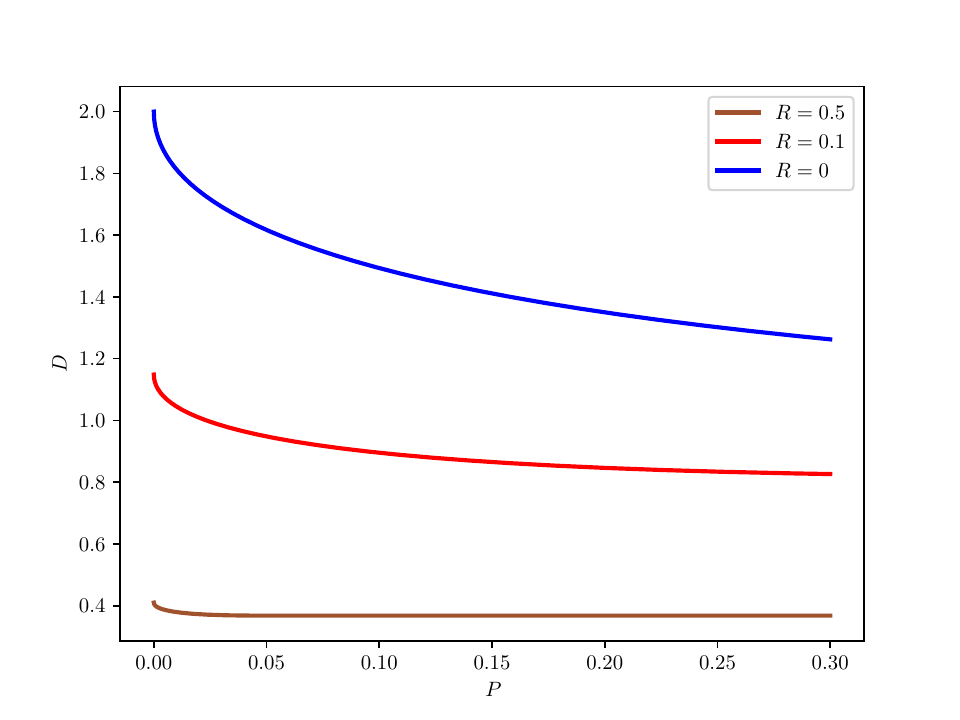}} \caption{Plots of $D(R,\infty,P|\mathcal{N}(0,1))$ with $R=0, 0.1, 0.5$.}
	\label{fig:Corollary2moreoverDP} 
\end{figure}

\begin{corollary}\label{cor:Gaussiantight}
	For the case  $\Delta(x,\hat{x})=(x-\hat{x})^2$ and $\phi(p_X,p_{\hat{X}})=\phi_{KL}(p_{\hat{X}}\|p_X)$, we have
	\begin{align}
		&\overline{D}(R,R_c,P|\mathcal{N}(\mu_X,\sigma^2_X))\nonumber\\
		&=\underline{D}(R,R_c,P|\mathcal{N}(\mu_X,\sigma^2_X))\nonumber\\
		&=\sigma^2_X+\sigma^2(P)-2\sigma_X\sigma(P)\xi(R,R_c)\label{eq:coincide}
	\end{align}
	if $\sigma(P)\geq\sigma(R,R_c,P)$, where
	\begin{align}
		\sigma(R,R_c,P):=\begin{cases}
			0&\mbox{if }R=0,\\
			\frac{\sqrt{1+4\beta_1\beta_2}-1}{2\beta_1}&\mbox{if }R>0
		\end{cases}\label{eq:sigma}
	\end{align}
with
\begin{align*}
\beta_1&=\frac{\sqrt{1-e^{-2R}}e^{-2(R+R_c)}}{\sigma_X\sqrt{1-e^{-2(R+R_c)}}},\\
\beta_2&=\sigma_X\sqrt{(1-e^{-2R})(1-e^{-2(R+R_c+P)})}\\
&\quad+\frac{\sigma_X\sqrt{1-e^{-2R}}e^{-2(R+R_c)}}{\sqrt{1-e^{-2(R+R_c)}}}.
\end{align*}
\end{corollary}
\begin{IEEEproof}
	See Appendix \ref{app:Gaussiantight}.
\end{IEEEproof}
\begin{remark}
	It can be seen from the proof of Corollary \ref{cor:Gaussiantight} that $\sigma(P)\geq\sigma(R,R_c,P)$ is  just a sufficient condition for (\ref{eq:coincide}) to hold.
	On the other hand, as shown in Appendix \ref{app:Gaussiantight}, when $R>0$, a necessary condition for  (\ref{eq:coincide}) to hold is 
	$\sigma(P)\geq\varsigma(R,R_C)$, where
	\begin{align}
		&\varsigma(R,R_c)\nonumber\\
		&:=\frac{\sigma_X\sqrt{1-e^{-2(R+R_c)}+4(1-e^{-2R})e^{-2(R+R_c)}}}{2\sqrt{1-e^{-2R}}e^{-2(R+R_c)}}\nonumber\\
		&\quad-\frac{\sigma_X\sqrt{1-e^{-2(R+R_c)}}}{2\sqrt{1-e^{-2R}}e^{-2(R+R_c)}}.\label{eq:varsigma}
	\end{align}
It is clear that for $R>0$, we must have 
\begin{align*}
	\sigma(R,R_c,P)\geq\varsigma(R,R_c).
\end{align*}	
Moreover, as shown in Appendix \ref{app:Gaussiantight}, for $R>0$,
	\begin{align}
		\sigma(R,R_c,P)\leq\varsigma'(R,R_c),\label{eq:sigma'}
	\end{align}
where
\begin{align}
	\varsigma'(R,R_c):=\frac{\sqrt{1+4\beta_1\beta'_2}-1}{2\beta_1}\label{eq:varsigma'}
\end{align}
with
\begin{align*}
	\beta'_2:=\sigma_X\sqrt{1-e^{-2R}}\left(1+\frac{e^{-2(R+R_c)}}{\sqrt{1-e^{-2(R+R_c)}}}\right).
\end{align*}
Since $\varsigma'(R,R_c)<\sigma_X$, it follows that $\sigma(P)\geq\varsigma'(R,R_c)$ and consequently $\sigma(P)\geq\sigma(R,R_c,P)$ when $P$ is sufficiently close to $0$, confirming the phenomenon in Fig. \ref{fig:Theorem3PD}. Also note that for $R_c>0$, 
\begin{align*}
	\lim\limits_{R\rightarrow 0}\varsigma'(R,R_c)=0.
\end{align*}
Therefore, given $R_c>0$ and $P<\infty$, we have $\sigma(P)\geq\varsigma'(R,R_c)\geq\sigma(R,R_c,P)$ when $R$ is sufficiently close to $0$, confirming the phenomenon in Fig. \ref{fig:Theorem3RD}. However, the behavior of $\varsigma'(R,R_c)$ is quite different at $R_c=0$. Indeed, 
\begin{align*}
	\varsigma'(R,0)=\frac{\sigma_X(\sqrt{1+4(\sqrt{1-e^{-2R}}+e^{-2R})e^{-2R}}-1)}{2e^{-2R}},
\end{align*}
which gives
\begin{align*}
	\lim\limits_{R\rightarrow 0}\varsigma'(R,0)=\frac{\sigma_X(\sqrt{5}-1)}{2}.
\end{align*}
It turns out that $\varsigma(R,R_c)$ also has the above properties. Specifically, for $R_c>0$, 
\begin{align*}
	\lim\limits_{R\rightarrow 0}\varsigma(R,R_c)=0;
\end{align*}
in contrast,
\begin{align*}
	\varsigma(R,0)=\frac{\sigma_X(\sqrt{1+4e^{-2R}}-1)}{2e^{-2R}}
\end{align*}
and consequently 
\begin{align*}
	\lim\limits_{R\rightarrow 0}\varsigma(R,0)=\frac{\sigma_X(\sqrt{5}-1)}{2}.
\end{align*}
Note that $\sigma(R,R_c,P)$ must share the same properties as it is bounded between $\varsigma(R,R_c)$ and $\varsigma'(R,R_c)$ when $R>0$.


For the extreme case $P=0$, we have $\sigma(P)=\sigma_X$. The condition $\sigma(P)\geq \sigma(R,R_c,P)$ is trivially satisfied and
\begin{align*}
	&\overline{D}(R,R_c,0|\mathcal{N}(\mu_X,\sigma^2_X))\\
	&=\underline{D}(R,R_c,0|\mathcal{N}(\mu_X,\sigma^2_X))\\
	&=2\sigma^2_X-2\sigma^2_X\xi(R,R_c).
\end{align*} 
As a consequence, we have a complete characterization of $D(R,R_c,0|\mathcal{N}(\mu_X,\sigma^2_X))$, recovering \cite[Proposition 1]{Wagner22}. Fig. \ref{fig:Corollary3Case3DR_rc1} shows the plots of $D(R,R_c,0|\mathcal{N}(0,1))$ with $R_c=0, 1, \infty$. It can be seen that a small amount of common randomness is able to achieve almost the same effect on the distortion-rate tradeoff as unlimited common randomness.

Another extreme case of interest is $R_c=\infty$. We have
\begin{align*}
	&\underline{D}(R,\infty,P|\mathcal{N}(\mu_X,\sigma^2_X))\\
	&=\min\limits_{\sigma_{\hat{X}}\in[\sigma(P),\sigma_X]}\sigma^2_X+\sigma^2_{\hat{X}}-2\sigma_X\sigma_{\hat{X}}\sqrt{1-e^{-2R}}\\
	&=\begin{cases}
		\sigma^2_Xe^{-2R}&\hspace{-1.0in}\mbox{if }\sigma(P)\leq\sigma_X\sqrt{1-e^{-2R}},\\
		\sigma^2_X+\sigma^2(P)-2\sigma_X\sigma(P)\sqrt{1-e^{-2R}}\\&\hspace{-1.0in}\mbox{if }\sigma(P)>\sigma_X\sqrt{1-e^{-2R}},
	\end{cases}
\end{align*}
which coincides with $\overline{D}(R,\infty,P|\mathcal{N}(\mu_X,\sigma^2_X))$.
This results in a complete characterization of $D(R,\infty,P|\mathcal{N}(\mu_X,\sigma^2_X))$,
recovering \cite[Theorem 1]{ZQCK21} (see also \cite{SPCYK23,QSCKYSGT23} for various extensions). Fig. \ref{fig:Corollary2case1DR} shows the plots of $D(R,\infty,P|\mathcal{N}(0,1))$ with $P=0, 0.1, \infty$.
It can be seen that the perception constraint affects the distortion-rate tradeoff only when $R$ is below a certain threshold\footnote{It can be inferred from the condition $\sigma(P)>\sigma_X\sqrt{1-e^{-2R}}$ that with unlimited common randomness,
the perception constraint affects the distortion-rate tradeoff only when $R<\frac{1}{2}\log\frac{\sigma^2_X}{\sigma^2_X-\sigma^2(P)}$.}. 
To investigate the distortion-perception tradeoff at a given rate, 
we plot $D(R,\infty,P|\mathcal{N}(0,1))$ with $R=0, 0.1, 0.5$ in Fig. \ref{fig:Corollary2moreoverDP}.
One can readily see that this tradeoff is most visible when $R$ is small.
\end{remark}



\begin{figure}[htbp]
	\centerline{\includegraphics[width=9cm]{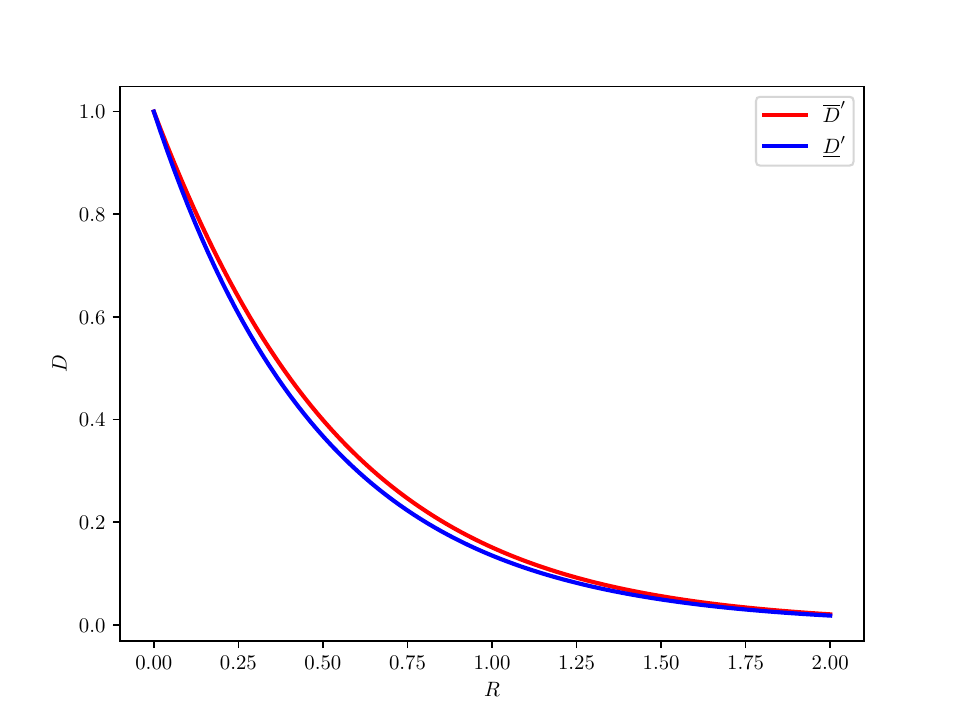}} \caption{Plots of $\overline{D}'(R,1,1|\mathcal{N}(0,1))$ and $\underline{D}'(R,1,1|\mathcal{N}(0,1))$.}
	\label{fig:theroem4Case2RD} 
\end{figure}

\begin{figure}[htbp]
	\centerline{\includegraphics[width=9cm]{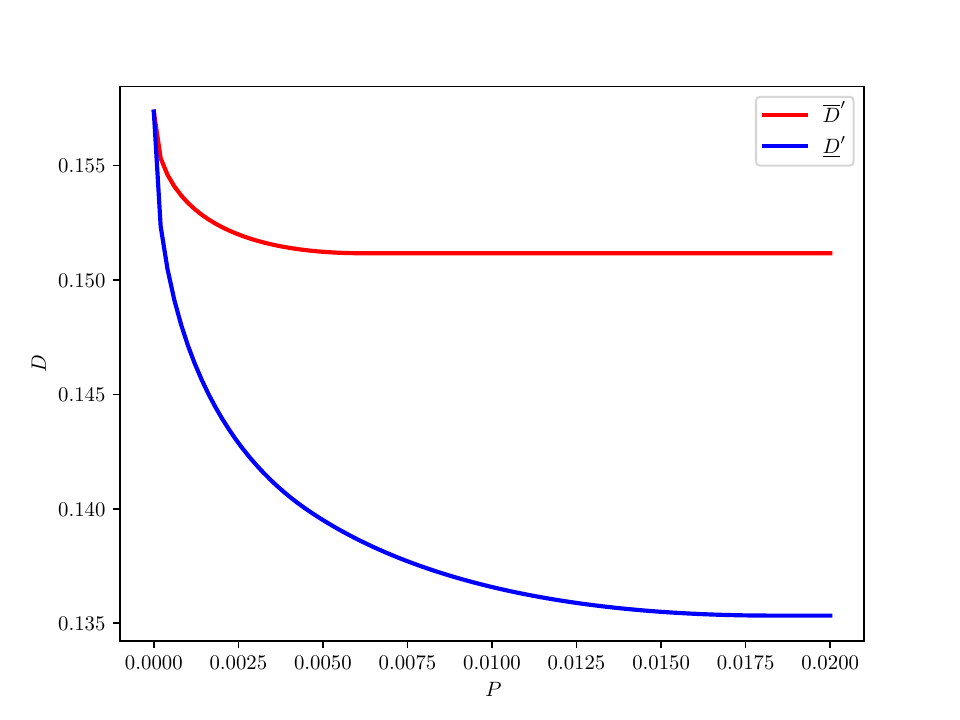}} \caption{Plots of $\overline{D}'(1,1,P|\mathcal{N}(0,1))$ and $\underline{D}'(1,1,P|\mathcal{N}(0,1))$.}
	\label{fig:theroem4Case2PD} 
\end{figure}

\begin{theorem}\label{thm:GaussianW2}
For the case $\Delta(x,\hat{x})=(x-\hat{x})^2$ and $\phi(p_X,p_{\hat{X}})=W^2_2(p_X,p_{\hat{X}})$, we have
\begin{align}
	&D(R,R_c,P|\mathcal{N}(\mu_X,\sigma^2_X))\leq\overline{D}'(R,R_c,P|\mathcal{N}(\mu_X,\sigma^2_X)),\label{eq:W2upper}\\
	&D(R,R_c,P|\mathcal{N}(\mu_X,\sigma^2_X))\geq\underline{D}'(R,R_c,P|\mathcal{N}(\mu_X,\sigma^2_X)),\label{eq:W2lower}
\end{align}
where
\begin{align*}
	&\overline{D}'(R,R_c,P|\mathcal{N}(\mu_X,\sigma^2_X))\\
	&:=\begin{cases}
		\sigma^2_X-\sigma^2_X\xi^2(R,R_c)&\hspace{-1.5in}\mbox{if }\sigma_X-\sqrt{P}\leq\sigma_X\xi(R,R_c),\\
		\sigma^2_X+(\sigma_X-\sqrt{P})^2-2\sigma_X(\sigma_X-\sqrt{P})\xi(R,R_c)\\
		&\hspace{-1.5in}\mbox{if }\sigma_X-\sqrt{P}>\sigma_X\xi(R,R_c),
	\end{cases}\\
	&\underline{D}'(R,R_c,P|\mathcal{N}(\mu_X,\sigma^2_X)):=\min\limits_{\sigma_{\hat{X}}\in[(\sigma_X-\sqrt{P})_+,\sigma_X]}\sigma^2_X+\sigma^2_{\hat{X}}\\
	&\hspace{0.35in}-2\sigma_X\sqrt{(1-e^{-2R})(\sigma^2_{\hat{X}}-(\sigma_Xe^{-(R+R_c)}-\sqrt{P})^2_+)}.
\end{align*}
\end{theorem}
\begin{IEEEproof}
	See Appendix \ref{app:GaussianW2}.
\end{IEEEproof}
\begin{remark}
	As shown in Appendix \ref{app:GaussianW2}, the minimization problem that defines $\underline{D}'(R,R_c,P|\mathcal{N}(\mu_X,\sigma^2_X))$ can be solved explicitly. Specifically, we have	
	\begin{align*}
		\underline{D}'(0,R_c,P|\mathcal{N}(\mu_X,\sigma^2_X))=\sigma^2_X+(\sigma_X-\sqrt{P})^2_+,
	\end{align*}
and for $R>0$,	
	\begin{align*}
		&\underline{D}'(R,R_c,P|\mathcal{N}(\mu_X,\sigma^2_X))\\
		&=\begin{cases}
			\sigma^2_Xe^{-2R}&\hspace{-2.2in}\mbox{if }\frac{\sqrt{P}}{\sigma_X}\geq(1-\sqrt{1-e^{-2R}})\vee e^{-(R+R_c)},\\
			\sigma^2_X+(\sigma_X-\sqrt{P})^2-2\sigma_X(\sigma_X-\sqrt{P})\sqrt{1-e^{-2R}}\\
			&\hspace{-2.2in}\mbox{if }\frac{\sqrt{P}}{\sigma_X}\in[e^{-(R+R_c)},1-\sqrt{1-e^{-2R}}),\\
			\sigma^2_Xe^{-2R}+(\sigma_Xe^{-(R+R_c)}-\sqrt{P})^2\\
			&\hspace{-2.2in}\mbox{if }\frac{\sqrt{P}}{\sigma_X}\in[\nu(R,R_c),e^{-(R+R_c)}),\\
			\sigma^2_X+(\sigma_X-\sqrt{P})^2-2\sigma^2_X\sqrt{(1-e^{-2R})}\\
			\times\sqrt{(1-e^{-(R+R_c)})(1+e^{-(R+R_c)}-\frac{2\sqrt{P}}{\sigma_X})}\\
			&\hspace{-2.2in}\mbox{if }\frac{\sqrt{P}}{\sigma_X}<\nu(R,R_c)\wedge e^{-(R+R_c)},
		\end{cases}
	\end{align*}
where
\begin{align}
	\nu(R,R_c):=\frac{e^{-2R}-e^{-2(R+R_c)}}{2-2e^{-(R+R_c)}}.\label{eq:nu}
\end{align}
Note that $e^{-(R+R_c)}<1-\sqrt{1-e^{-2R}}$ if and only if
\begin{align*}
	R<\log\frac{1+e^{-2R_c}}{2e^{-R_c}}
\end{align*}
while $\nu(R,R_c)<e^{-(R+R_c)}$ if and only if
\begin{align*}
	R>\log\frac{1+e^{-2R_c}}{2e^{-R_c}}.
\end{align*}
Therefore, the two intervals $[e^{-(R+R_c)},1-\sqrt{1-e^{-2R}})$ and $[\nu(R,R_c),e^{-(R+R_c)})$ cannot be non-empty simultaneously.
	
	Fig. \ref{fig:theroem4Case2RD} shows the plots of $\overline{D}'(R,1,1|\mathcal{N}(0,1))$ and $\underline{D}'(R,1,1|\mathcal{N}(0,1))$. It can be seen that they are quite close to each other. However, different from the case $\phi(p_X,p_{\hat{X}})=\phi_{KL}(p_{\hat{X}}\|p_X)$, in general the two bounds do not match exactly even in the low rate region. Similarly, Fig. \ref{fig:theroem4Case2PD} shows that $\overline{D}'(1,1,P|\mathcal{N}(0,1))$ and $\underline{D}'(1,1,P|\mathcal{N}(0,1))$ do not coincide even when $P$ is small.

	However, there are two exceptions: 1) $P=0$ and 2) $R_c=\infty$. Specifically, we have	
	\begin{align*}
		&\overline{D}'(R,R_c,0|\mathcal{N}(\mu_X,\sigma^2_X))\\
		&=\underline{D}'(R,R_c,0|\mathcal{N}(\mu_X,\sigma^2_X))\\
		&=2\sigma^2_X-2\sigma^2_X\xi(R,R_c)
	\end{align*}	
		and
		\begin{align*}
		&\overline{D}'(R,\infty,P|\mathcal{N}(\mu_X,\sigma^2_X))\\
		&=\underline{D}'(R,\infty,P|\mathcal{N}(\mu_X,\sigma^2_X))\\
		&=\begin{cases}
			\sigma^2_Xe^{-2R}			&\hspace{-1.7in}\mbox{if }\sigma_X-\sqrt{P}\leq\sigma_X\sqrt{1-e^{-2R}},\\
			\sigma^2_X+(\sigma_X-\sqrt{P})^2-2\sigma_X(\sigma_X-\sqrt{P})\sqrt{1-e^{-2R}}\\
			&\hspace{-1.7in}\mbox{if }\sigma_X-\sqrt{P}>\sigma_X\sqrt{1-e^{-2R}}.
		\end{cases}
	\end{align*}
So $D(R,R_c,0|\mathcal{N}(\mu_X,\sigma^2_X))$ and $D(R,\infty,P|\mathcal{N}(\mu_X,\sigma^2_X))$ are completely characterized.

	It is worth noting that
	\begin{align*} &\left.D(R,R_c,0|\mathcal{N}(\mu_X,\sigma^2_X))\right|_{\phi=\phi_{KL}}\\
		&=\left.D(R,R_c,0|\mathcal{N}(\mu_X,\sigma^2_X))\right|_{\phi=W^2_2}.
	\end{align*}	
		 This is not surprising because regardless of the choice of $\phi$, the constraint $\phi(p_X,p_{\hat{X}})\leq 0$ is equivalent to setting $p_{\hat{X}}=p_X$. It can also be seen that
		 \begin{align*}
		 &\left.D(R,\infty,P|\mathcal{N}(\mu_X,\sigma^2_X))\right|_{\phi=\phi_{KL}}\\
		 &=\left.D(R,\infty,(\sigma_X-\sigma(P))^2|\mathcal{N}(\mu_X,\sigma^2_X))\right|_{\phi=W^2_2} 
		 \end{align*}
		 as pointed out in \cite[Theorem 1]{ZQCK21}.
		 More generally, we have\footnote{Note that $\overline{D}(R,R_c,P|\mathcal{N}(\mu_X,\sigma^2_X))$ and $\overline{D}'(R,R_c,P|\mathcal{N}(\mu_X,\sigma^2_X))$ coincide respectively with $\left.D(R,\infty,P|\mathcal{N}(\mu_X,\sigma^2_X))\right|_{\phi=\phi_{KL}}$ and $\left.D(R,\infty,P|\mathcal{N}(\mu_X,\sigma^2_X))\right|_{\phi=W^2_2}$ when $R_c=\infty$.}
		 \begin{align*}
		 	&\overline{D}(R,R_c,P|\mathcal{N}(\mu_X,\sigma^2_X))\\
		 	&=\overline{D}'(R,R_c,(\sigma_X-\sigma(P))^2|\mathcal{N}(\mu_X,\sigma^2_X)). 
	 	\end{align*} 
 	This correspondence is a consequence of the fact that both upper bounds are established by restricting $p_{\hat{X}}$ to be Gaussian with $\mu_{\hat{X}}=\mu_{X}$ and $\sigma_{\hat{X}}\leq\sigma_X$, which induces a one-to-one map between $\phi_{KL}(p_{\hat{X}}\|\mathcal{N}(\mu_X,\sigma^2_{X}))$ and $W^2_2(\mathcal{N}(\mu_X,\sigma^2_{X}),p_{\hat{X}})$.
\end{remark}


\section{Conclusion}\label{sec:conclusion}

By exploring the connection between output-constrained lossy source coding and rate-distotion-perception coding, 
we have estabilished upper and lower bounds on the fundamental rate-distortion-perception tradeoff with limited common randomness for the quadratic Gaussian case when the perception measure is given by Kullback-Leibler divergence or squared quadratic Wasserstein distance. It is of considerable interest to further investigate the tightness of our bounds.

We will end this paper with a brief comment on the formulation of perception constraint. Note that the reconstructed symbols are required to be i.i.d. in our work. As such, it suffices to adopt marginal-distribution-based perception constraint (\ref{eq:marginal_perception}) to control the sequence-level distributional difference between the source and reconstruction. Without the i.i.d. requirement on the reconstructed symbols\footnote{The source symbols are still assumed to be i.i.d.}, one may replace (\ref{eq:marginal_perception}) with joint-distribution-based perception constraint 
\begin{align}
\frac{1}{n}\phi(p_{X^n},p_{\hat{X}^n})\leq P\label{eq:joint_perception}
\end{align}
to enforce the sequence-level distributional similarity. Assuming $\phi$ is tensorizable\footnote{This is the case for Kullback-Leibler divergence and squared quadratic Wasserstein distance.} in the sense that
\begin{align*} 
	\phi(\otimes_{t=1}^n p_{X_t},\otimes_{t=1}^n p_{\hat{X}_t})=\sum_{t=1}^n\phi(p_{X_t},p_{\hat{X}_t}),
	\end{align*} 
joint-distribution-based perception constraint
(\ref{eq:joint_perception}) without the  i.i.d. requirement is a relaxed version of marginal-distribution-based perception constraint (\ref{eq:marginal_perception}) with the i.i.d. requirement. Characterizing the impact of this relaxation on the fundamental rate-distortion-perception tradeoff is left for future work.



%
%
%
%


%

\appendices
\section{Proof of Theorem \ref{thm:outputconstrained}}\label{app:outputconstrained}

As the single-letter characterization in (\ref{eq:expression1})--(\ref{eq:constraintb}) is implied by \cite[Theorem 2]{Wagner22} and  \cite[Theorem 1]{SLY15J2}, it suffices to prove the specialized form in (\ref{eq:expression2})--(\ref{eq:constraint4}). Note that $(p_X,p_{\hat{X}},\Delta)$ is uniformly integrable when $\mathbb{E}[X^2]<\infty$, $\mathbb{E}[\hat{X}^2]<\infty$, and $\Delta(x,\hat{x})=(x-\hat{x})^2$ \cite{CYWSGT22}.

For any $p_{XU\hat{X}}\in\Omega(p_X,p_{\hat{X}})$ satisfying (\ref{eq:constrainta}) and (\ref{eq:constraintb}), let
\begin{align}
	&Y:=\mathbb{E}[X|U],\label{eq:1}\\
	&\hat{Y}:=\mathbb{E}[\hat{X}|U].\label{eq:2}
\end{align}
By the data processing inequality \cite[Theorem 2.8.1]{CT91}, 
\begin{align}
	&I(X;Y)\leq I(X;U)\leq R,\label{eq:3}\\
	&I(\hat{X};\hat{Y})\leq I(\hat{X};U)\leq R+R_c.\label{eq:4}
\end{align}
It can be verified that 
\begin{align}
	&\mathbb{E}[(X-\hat{X})^2]\nonumber\\
	&=\mathbb{E}[((X-Y)+(\hat{X}-\hat{Y})+(Y-\hat{Y}))^2]\nonumber\\
	&=\mathbb{E}[(X-Y)^2]+\mathbb{E}[(\hat{X}-\hat{Y})^2]+\mathbb{E}[(Y-\hat{Y})^2]\nonumber\\
	&\quad+2\mathbb{E}[(X-Y)(\hat{X}-\hat{Y})]+2\mathbb{E}[(X-Y)(Y-\hat{Y})]\nonumber\\
	&\quad+2\mathbb{E}[(\hat{X}-\hat{Y})(Y-\hat{Y})].\label{eq:subed}
\end{align}
We have	
\begin{align}
	\mathbb{E}[(X-Y)(\hat{X}-\hat{Y})]&=\mathbb{E}[\mathbb{E}[(X-Y)(\hat{X}-\hat{Y})|U]]\nonumber\\
	&\stackrel{(a)}{=}\mathbb{E}[\mathbb{E}[X-Y|U]\mathbb{E}[\hat{X}-\hat{Y}|U]]\nonumber\\
	&=0,\label{eq:zero1}
\end{align}
where ($a$)	is due to the conditional independence of $X-Y$ and $\hat{X}-\hat{Y}$ given $U$.  Similarly, 	
\begin{align}
	&\mathbb{E}[(X-Y)(Y-\hat{Y})]=0,\label{eq:zero2}\\
	&\mathbb{E}[(\hat{X}-\hat{Y})(Y-\hat{Y})]=0.\label{eq:zero3}
\end{align}
Moreover, 
\begin{align}
	\mathbb{E}[(Y-\hat{Y})^2]\geq W^2_2(p_Y,p_{\hat{Y}}).\label{eq:W2bound}
\end{align}
Substituting (\ref{eq:zero1})--(\ref{eq:W2bound}) into (\ref{eq:subed}) gives
\begin{align*}
	\mathbb{E}[(X-Y)^2]\geq\mathbb{E}[(X-Y)^2]+\mathbb{E}[(\hat{X}-\hat{Y})^2]+W^2_2(p_{Y},p_{\hat{Y}}).
\end{align*}
In light of (\ref{eq:1})--(\ref{eq:4}), the constructed $Y$ and $\hat{Y}$ satisfy (\ref{eq:constraint1})-(\ref{eq:constraint4}). Therefore, it follows by (\ref{eq:expression1}) that
\begin{align*}
	&D(R,R_c|p_X,p_{\hat{X}})\nonumber\\
	&\geq\inf\limits_{p_{Y|X},p_{\hat{Y}|\hat{X}}}\mathbb{E}[(X-Y)^2]+\mathbb{E}[(\hat{X}-\hat{Y})^2]+W^2_2(p_{Y},p_{\hat{Y}})
\end{align*}
subject to (\ref{eq:constraint1})--(\ref{eq:constraint4}).

Now we shall prove that this lower bound is tight. 
Let $Y$ and $\hat{Y}$ be jointly distributed with $X$ and $\hat{X}$, respectively, such that (\ref{eq:constraint1})--(\ref{eq:constraint4}) are satisfied. Given $p_{XY}$ and $p_{\hat{X}\hat{Y}}$, construct $p_{XY\hat{Y}\hat{X}}$ with $X\leftrightarrow Y\leftrightarrow\hat{Y}\leftrightarrow\hat{X}$ forming a Markov chain and\footnote{It is known \cite[Theorem 1.3]{Villani03} that there exists a coupling of $p_Y$ and $p_{\hat{Y}}$ for which (\ref{eq:W22}) holds. In other words, the infimum in (\ref{eq:inf}) can be attained.}
\begin{align}
	\mathbb{E}[(Y-\hat{Y})^2]=W^2_2(p_Y,p_{\hat{Y}}).\label{eq:W22}
\end{align}
Note that (\ref{eq:subed}) continues to hold for the constructed $p_{XY\hat{Y}\hat{X}}$, i.e.,
\begin{align}
	&\mathbb{E}[(X-\hat{X})^2]\nonumber\\
	&=\mathbb{E}[(X-Y)^2]+\mathbb{E}[(\hat{X}-\hat{Y})^2]+\mathbb{E}[(Y-\hat{Y})^2]\nonumber\\
	&\quad+2\mathbb{E}[(X-Y)(\hat{X}-\hat{Y})]+2\mathbb{E}[(X-Y)(Y-\hat{Y})]\nonumber\\
	&\quad+2\mathbb{E}[(\hat{X}-\hat{Y})(Y-\hat{Y})].\label{eq:tbs}
\end{align}
We have
\begin{align}
	\mathbb{E}[(X-Y)(\hat{X}-\hat{Y})]&=\mathbb{E}[\mathbb{E}[(X-Y)(\hat{X}-\hat{Y})|Y]]\nonumber\\
	&\stackrel{(b)}{=}\mathbb{E}[\mathbb{E}[X-Y|Y]\mathbb{E}[\hat{X}-\hat{Y}|Y]]\nonumber\\
	&=0,\label{eq:0_1}
\end{align}
where ($b$) is due to the conditional independence of $X-Y$ and $\hat{X}-\hat{Y}$ given $Y$.
Similarly, 
\begin{align}
	&\mathbb{E}[(X-Y)(Y-\hat{Y})]=0,\label{eq:0_2}\\
	&\mathbb{E}[(\hat{X}-\hat{Y})(Y-\hat{Y})]=0.\label{eq:0_3}
\end{align}
Substituting (\ref{eq:W22}) and (\ref{eq:0_1})--(\ref{eq:0_3}) into (\ref{eq:tbs}) gives
\begin{align*}
	\mathbb{E}[(X-\hat{X})^2]=\mathbb{E}[(X-Y)^2]+\mathbb{E}[(\hat{X}-\hat{Y})^2]+W^2_2(p_Y,p_{\hat{Y}}).
\end{align*}
Let $U:=Y$. It is clear that $p_{XU\hat{X}}\in\Omega(p_X,p_{\hat{X}})$; moreover, (\ref{eq:constrainta}) and (\ref{eq:constraintb}) are satisfied since
\begin{align*}
	&I(X;U)=I(X;Y)\leq R,\\
	&I(\hat{X};U)\stackrel{(c)}{\leq} I(\hat{X},\hat{Y})\leq R+R_c,
\end{align*}
where ($c$) is due to the data processing inequality \cite[Theorem 2.8.1]{CT91}. Therefore, it follows by (\ref{eq:expression1}) that
\begin{align*}
	&D(R,R_c|p_X,p_{\hat{X}})\nonumber\\
	&\leq\inf\limits_{p_{Y|X},p_{\hat{Y}|\hat{X}}}\mathbb{E}[(X-Y)^2]+\mathbb{E}[(\hat{X}-\hat{Y})^2]+W^2_2(p_{Y},p_{\hat{Y}})
\end{align*}
subject to (\ref{eq:constraint1})--(\ref{eq:constraint4}). This completes the proof of Theorem \ref{thm:outputconstrained}.

\section{Proof of Corollary \ref{cor:generallowerbound}}\label{app:generallowerbound}

It is clear that $Y^*$ and $\hat{Y}^*$ satisfy (\ref{eq:constraint1})--(\ref{eq:constraint4}). Therefore, in light of Theorem \ref{thm:outputconstrained},
\begin{align*}
	&D(R,R_c|p_X,p_{\hat{X}})\\
	&\leq\mathbb{E}[(X-Y^*)^2]+\mathbb{E}[(\hat{X}-\hat{Y}^*)]+W_2^2(p_{Y^*},p_{\hat{Y}^*})\\
	&=D(R|p_X)+D(R+R_c|p_{\hat{X}})+W_2^2(p_{Y^*},p_{\hat{Y}^*}),
\end{align*}
which proves $D(R,R_c|p_X,p_{\hat{X}})\leq\overline{D}(R,R_c|p_X,p_{\hat{X}})$.

Now let $Y$ and $\hat{Y}$ be jointly distributed with $X$ and $\hat{X}$, respectively, such that (\ref{eq:constraint1})--(\ref{eq:constraint4}) are satisfied. We have
\begin{align}
	\mu_{Y}\stackrel{(a)}{=}\mathbb{E}[\mathbb{E}[X|Y]]=\mu_X\label{eq:mean1}
\end{align}
and
\begin{align}
	\mathbb{E}[(X-Y)^2]&=\mathbb{E}[((X-\mu_Y)-(Y-\mu_Y))^2]\nonumber\\
	&=\sigma^2_X+\sigma^2_Y-2\mathbb{E}[(X-\mu_Y)(Y-\mu_Y)]\nonumber\\
	&=\sigma^2_X+\sigma^2_Y-2\mathbb{E}[\mathbb{E}[(X-\mu_Y)(Y-\mu_Y)|Y]]\nonumber\\
	&=\sigma^2_X+\sigma^2_Y-2\mathbb{E}[(Y-\mu_Y)\mathbb{E}[(X-\mu_Y)|Y]]\nonumber\\
	&\stackrel{(b)}{=}\sigma^2_X+\sigma^2_Y-2\mathbb{E}[(Y-\mu_Y)^2]\nonumber\\
	&=\sigma^2_X-\sigma^2_Y,\label{eq:sigmadifference}
\end{align}
where ($a$) and ($b$) are due to (\ref{eq:constraint1}). Similarly, 
\begin{align}
	&\mu_{\hat{Y}}=\mu_{\hat{X}},\label{eq:mean2}\\
	&\mathbb{E}[(\hat{X}-\hat{Y})^2]=\sigma^2_{\hat{X}}-\sigma^2_{\hat{Y}}.\label{eq:sigmadifference2}	
\end{align}
In light of Proposition \ref{prop:W2},
\begin{align}
	W^2_2(p_Y,p_{\hat{Y}})&\geq(\mu_Y-\mu_{\hat{Y}})^2+(\sigma_Y-\sigma_{\hat{Y}})^2\nonumber\\
	&=(\mu_X-\mu_{\hat{X}})+\sigma^2_Y+\sigma^2_{\hat{Y}}-2\sigma_Y\sigma_{\hat{Y}}.\label{eq:W2}
\end{align}
Combining (\ref{eq:sigmadifference}), (\ref{eq:sigmadifference2}), and (\ref{eq:W2}) gives
\begin{align}
	&\mathbb{E}[(X-Y)^2]+\mathbb{E}[(\hat{X}-\hat{Y})^2]+W^2_2(p_Y,p_{\hat{Y}})\nonumber\\
	&\geq(\mu_X-\mu_{\hat{X}})^2+\sigma^2_X+\sigma^2_{\hat{X}}-2\sigma_Y\sigma_{\hat{Y}}.\label{eq:sub}
\end{align}
Note that (\ref{eq:constraint3}) implies $\mathbb{E}[(X-Y)^2]\geq D(R|p_X)$, which, together with (\ref{eq:sigmadifference}), further implies
\begin{align}
	\sigma_Y\leq\sqrt{\sigma^2_X-D(R|p_X)}.\label{eq:sub1}
\end{align}
Similarly, we have
\begin{align}
	\sigma_{\hat{Y}}\leq\sqrt{\sigma^2_{\hat{X}}-D(R+R_c|p_{\hat{X}})}.\label{eq:sub2}
\end{align}
Substituting (\ref{eq:sub1}) and (\ref{eq:sub2}) into (\ref{eq:sub}) yields 
\begin{align}
	&D(R,R_c|p_X,p_{\hat{X}})
	\nonumber\\
	&\geq(\mu_X-\mu_{\hat{X}})^2+\sigma^2_X+\sigma^2_{\hat{X}}\nonumber\\
	&\quad-2\sqrt{(\sigma_X^2-D(R|p_X))(\sigma^2_{\hat{X}}-D(R+R_c|p_{\hat{X}}))}. \label{eq:yield1}
\end{align}
It can be verified that
\begin{align}
	&\mathbb{E}[(X-Y^*)^2]+\mathbb{E}[(\hat{X}-\hat{Y}^*)^2]+W^2_2(p_{Y^G},p_{\hat{Y}^G})\nonumber\\
	&\stackrel{(c)}=\mathbb{E}[(X-Y^*)^2]+\mathbb{E}[(\hat{X}-\hat{Y}^*)^2]+(\mu_{Y^*}-\mu_{\hat{Y}^*})^2\nonumber\\
	&\quad+(\sigma_{Y^*}-\sigma_{\hat{Y}^*})^2\nonumber\\
	&\stackrel{(d)}{=}\mathbb{E}[(X-Y^*)^2]+\mathbb{E}[(\hat{X}-\hat{Y}^*)^2]+(\mu_{X}-\mu_{\hat{X}})^2\nonumber\\
	&\quad+(\sigma_{Y^*}-\sigma_{\hat{Y}^*})^2\nonumber\\
	&\stackrel{(e)}{=}(\mu_{X}-\mu_{\hat{X}})^2+\sigma^2_X+\sigma^2_{\hat{X}}-2\sigma_{Y^*}\sigma_{\hat{Y}^*}\nonumber\\
	&\stackrel{(f)}{=}(\mu_X-\mu_{\hat{X}})^2+\sigma^2_X+\sigma^2_{\hat{X}}\nonumber\\
	&\quad-2\sqrt{(\sigma_X^2-\mathbb{E}[(X-Y^*)^2])(\sigma^2_{\hat{X}}-\mathbb{E}[(\hat{X}-\hat{Y}^*)^2])},\label{eq:yield2}
\end{align}
where ($c$) is due to Proposition \ref{prop:W2} while ($d$)--($f$) is because
 (\ref{eq:mean1})--(\ref{eq:sigmadifference2}) also holds for $Y^*$ and $\hat{Y}^*$ as they satisfy (\ref{eq:constraint1})--(\ref{eq:constraint4}). 
Combining (\ref{eq:yield1}) and (\ref{eq:yield2}) as well as the fact that $\mathbb{E}[(X-Y^*)^2]=D(R|p_X)$ and $\mathbb{E}[(\hat{X}-\hat{Y}^*)^2]=D(R+R_c|p_{\hat{X}})$
 proves $D(R,R_c|p_X,p_{\hat{X}})\geq\underline{D}(R,R_c|p_X,p_{\hat{X}})$. 



\section{Proof of Theorem \ref{thm:Gaussianoutputconstrained}}\label{app:Gaussianoutputconstrained}

According to \cite[Theorem 13.3.2]{CT91}, 
\begin{align}
	&D(R|\mathcal{N}(\mu_X,\sigma^2_X))=\sigma^2_Xe^{-2R},\label{eq:s1}\\
	&D(R+R_c|\mathcal{N}(\mu_{\hat{X}},\sigma^2_{\hat{X}}))=\sigma^2_{\hat{X}}e^{-2(R+R_c)},\label{eq:s2}
\end{align}
and the outputs of the corresponding optimal test channels are normally distributed. As a consequence, $\overline{D}(R,R_c|\mathcal{N}(\mu_X,\sigma^2_X),\mathcal{N}(\mu_{\hat{X}},\sigma^2_{\hat{X}}))$ must coincide with $\underline{D}(R,R_c|\mathcal{N}(\mu_X,\sigma^2_X),\mathcal{N}(\mu_{\hat{X}},\sigma^2_{\hat{X}}))$. 
Now one can readily  show
\begin{align}
	&D(R,R_c|\mbox(\mu_X,\sigma^2_X),\mathcal{N}(\mu_{\hat{X}},\sigma^2_{\hat{X}}))\nonumber\\
	&=(\mu_X-\mu_{\hat{X}})^2+\sigma^2_X+\sigma^2_{\hat{X}}-2\sqrt{\sigma_X^2-D(R|\mathcal{N}(\mu_X,\sigma^2_X))}\nonumber\\
	&\quad\times\sqrt{\sigma^2_{\hat{X}}-D(R+R_c|\mathcal{N}(\mu_{\hat{X}},\sigma^2_{\hat{X}}))}\label{eq:s3}
	\end{align}
by invoking Corollary \ref{cor:generallowerbound} and (\ref{eq:remark}). 
Substituting (\ref{eq:s1}) and (\ref{eq:s2}) into (\ref{eq:s3}) gives	
\begin{align*}			&D(R,R_c|\mbox(\mu_X,\sigma^2_X),\mathcal{N}(\mu_{\hat{X}},\sigma^2_{\hat{X}}))\nonumber\\
	&=(\mu_X-\mu_{\hat{X}})^2+\sigma^2_X+\sigma^2_{\hat{X}}-2\sigma_X\sigma_{\hat{X}}\xi(R,R_c),
\end{align*}
which completes the proof of Theorem \ref{thm:Gaussianoutputconstrained}.

\section{Proof of Theorem \ref{thm:GaussianRDP}}\label{app:GaussianRDP}

\begin{lemma}\label{lem:suffKL}
	For the case $\Delta(x,\hat{x})=(x-\hat{x})^2$ and $\phi(p_X,p_{\hat{X}})=\phi(p_{\hat{X}}\|p_X)$, we have
	\begin{align*}
		D(R,R_c,P|\mathcal{N}(\mu_X,\sigma^2_X))=&\inf\limits_{p_{\hat{X}}}D(R,R_c|\mathcal{N}(\mu_X,\sigma^2_X),p_{\hat{X}})\\
		\mbox{subject to}&\quad \mu_{\hat{X}}=\mu_X,\\
		&\quad\sigma_{\hat{X}}\leq\sigma_X,\\
		&\quad\phi_{KL}(p_{\hat{X}}\|\mathcal{N}(\mu_X,\sigma^2_X))\leq P.
	\end{align*}
\end{lemma}
\begin{IEEEproof}
	We shall  show that there is no loss of optimality in replacing $p_{\hat{X}}$ by $p_{\hat{X}'}$ when $\sigma_{\hat{X}}>\sigma_X$, where
	$\hat{X}':=\frac{\sigma_X}{\sigma_{\hat{X}}}(\hat{X}-\mu_{\hat{X}})+\mu_X$.
	Clearly, if $X\leftrightarrow U\leftrightarrow\hat{X}$ form a Markov chain, then $X\leftrightarrow U\leftrightarrow\hat{X}'$ also form a Markov chain. Moreover, we have
	\begin{align*}
		I(\hat{X}';U)=I(\hat{X};U).
	\end{align*}
	It can be verified that
	\begin{align}
		&\mathbb{E}[(X-\hat{X}')^2]\nonumber\\
		&=\mathbb{E}\left[\left((X-\mu_X)-\frac{\sigma_X}{\sigma_{\hat{X}}}(\hat{X}-\mu_{\hat{X}})\right)^2\right]\nonumber\\
		&=2\sigma^2_X-\frac{2\sigma_X}{\sigma_{\hat{X}}}\mathbb{E}[(X-\mu_X)(\hat{X}-\mu_{\hat{X}})]\nonumber\\
		&\stackrel{(a)}{\leq}\sigma^2_X+\sigma^2_{\hat{X}}-2\mathbb{E}[(X-\mu_X)(\hat{X}-\mu_{\hat{X}})]\nonumber\\
		&\leq(\mu_X-\mu_{\hat{X}})^2+\sigma^2_X+\sigma^2_{\hat{X}}-2\mathbb{E}[(X-\mu_X)(\hat{X}-\mu_{\hat{X}})]\nonumber\\
		&=\mathbb{E}[(X-\hat{X})^2],\label{eq:variance}
	\end{align}
	where ($a$) is because  $c^2\sigma^2_{\hat{X}}-2c\mathbb{E}[(X-\mu_X)(\hat{X}-\mu_{\hat{X}})]$ is an increasing function of  $c$ for $c\in[\frac{\sigma_X}{\sigma_{\hat{X}}},1]$. In addition, 
	\begin{align*}
		&\phi_{KL}(p_{\hat{X}'}\|\mathcal{N}(\mu_X,\sigma^2_X))\\
		&=-h(\hat{X}')+\frac{1}{2}\log(2\pi\sigma^2_{X})+\frac{(\mu_X-\mu_{\hat{X}'})^2+\sigma^2_{\hat{X}'}}{2\sigma^2_X}\\
		&=-h(\hat{X})-\log\frac{\sigma_X}{\sigma_{\hat{X}}}+\frac{1}{2}\log(2\pi\sigma^2_X)+\frac{1}{2}\\
		&\stackrel{(b)}{\leq}-h(\hat{X})+\frac{1}{2}\log(2\pi\sigma^2_X)+\frac{\sigma^2_{\hat{X}}}{2\sigma^2_X}\\
		&\leq -h(\hat{X})+\frac{1}{2}\log(2\pi\sigma^2_X)+\frac{(\mu_X-\mu_{\hat{X}})^2+\sigma^2_{\hat{X}}}{2\sigma^2_X}\\
		&=\phi_{KL}(p_{\hat{X}}\|\mathcal{N}(\mu_X,\sigma^2_X)),
	\end{align*}
	where ($b$) is due to $\psi(\sigma_{\hat{X}})\geq 0$. This proves Lemma \ref{lem:suffKL}.
\end{IEEEproof}

In view of Lemma \ref{lem:suffKL}, it suffices to consider $p_{\hat{X}}$ with $\mu_{\hat{X}}=\mu_X$ and $\sigma_{\hat{X}}\leq\sigma_X$ for the purpose of computing $D(R,R_c,P|\mathcal{N}(\mu_X,\sigma^2_X))$.
Further restricting $p_{\hat{X}}$ to be Gaussian and invoking Theorem \ref{thm:Gaussianoutputconstrained} 	
yields the following upper bound on $D(R,R_c,P|\mathcal{N}(\mu_X,\sigma^2_X))$:
\begin{align}
	&D(R,R_c,P|\mathcal{N}(\mu_X,\sigma^2_X))\nonumber\\
	&\leq\min\limits_{\sigma_{\hat{X}}\in[\sigma(P),\sigma_X]}\sigma^2_X+\sigma^2_{\hat{X}}-2\sigma_X\sigma_{\hat{X}}\xi(R,R_c).\label{eq:inf2}
\end{align}
Clearly,
$\sigma^2_{\hat{X}}-2\sigma_X\sigma_{\hat{X}}\xi(R,R_c)$ is monotonically decreasing for $\sigma_{\hat{X}}\in[0,\sigma_X\xi(R,R_c)]$ and is monotonically increasing for $\sigma_{\hat{X}}\in[\sigma_X\xi(R,R_c),\infty)$. Therefore, the minimum  in (\ref{eq:inf2}) is attained at 
$\sigma_{\hat{X}}=\sigma_X\xi(R,R_c)$
when $\sigma(P)\leq\sigma_X\xi(R,R_c)$, and is attained at
$\sigma_{\hat{X}}=\sigma(P)$
when $\sigma(P)>\sigma_X\xi(R,R_c)$. This proves (\ref{eq:Gaussianupper}).

To prove (\ref{eq:Gaussianlower}), we need the following lemma.


\begin{lemma}\label{lem:KLD}
	For any $p_{\hat{X}}$ with $\phi_{KL}(p_{\hat{X}}\|\mathcal{N}(\mu_X,\sigma^2_X))\leq P$, we have
	\begin{align*}
		&D(R+R_c|p_{\hat{X}})\\
		&\geq\sigma^2_{\hat{X}}e^{-2(R+R_c+P-\phi_{KL}(\mathcal{N}(\mu_{\hat{X}},\sigma^2_{\hat{X}})\|\mathcal{N}(\mu_X,\sigma^2_{X})))}.
	\end{align*}
\end{lemma}	
\begin{IEEEproof}
	Note that $\phi_{KL}(p_{\hat{X}}\|\mathcal{N}(\mu_X,\sigma^2_X))\leq P$ implies
	\begin{align}
		&h(\hat{X})\nonumber\\
		&\geq\frac{1}{2}\log(2\pi e\sigma^2_{\hat{X}})+\phi_{KL}(\mathcal{N}(\mu_{\hat{X}},\sigma^2_{\hat{X}})\|\mathcal{N}(\mu_X,\sigma^2_{X}))-P.\label{eq:entropybound}
	\end{align}
	Combining (\ref{eq:Shannonlowerbound}) and (\ref{eq:entropybound}) proves Lemma \ref{lem:KLD}.	
\end{IEEEproof}

Now we are in a position to prove (\ref{eq:Gaussianlower}). Note that Proposition \ref{prop:KL}, together with the constraints $\mu_{\hat{X}}=\mu_X$, $\sigma_{\hat{X}}\leq\sigma_X$, and $\phi_{KL}(p_{\hat{X}}\|\mathcal{N}(\mu_X,\sigma^2_X))\leq P$, implies
\begin{align}
	\sigma_{\hat{X}}\in[\sigma(P),\sigma_X].\label{eq:psi}
\end{align}
According to \cite[Theorem 13.3.2]{CT91},
\begin{align}
	D(R|\mathcal{N}(\mu_X,\sigma^2_X))=\sigma^2_Xe^{-2R}.\label{eq:bound1}
\end{align}
Moreover, it follows by Lemma \ref{lem:KLD} with $\mu_{\hat{X}}=\mu_X$ that
\begin{align}
	D(R+R_c|p_{\hat{X}})\geq\sigma^2_{\hat{X}}e^{-2(R+R_c+P-\psi(\sigma_{\hat{X}}))}.\label{eq:bound2}
\end{align}
In view of (\ref{eq:psi})--(\ref{eq:bound2}), one can readily prove (\ref{eq:Gaussianlower}) by invoking Corollary \ref{cor:generallowerbound} and (\ref{eq:remark}).

\section{Proof of Corollary \ref{cor:Gaussiantight}}\label{app:Gaussiantight}

It is easy to see that (\ref{eq:coincide}) holds if and only if the minimum value of $\zeta(\sigma_{\hat{X}})$ for $\sigma_{\hat{X}}\in[\sigma(P),\sigma_X]$ is attained at $\sigma_{\hat{X}}=\sigma(P)$, where
\begin{align*}
	\zeta(\sigma_{\hat{X}}):=\sigma^2_X+\sigma^2_{\hat{X}}-2\sigma_X\sigma_{\hat{X}}\tau(\sigma_{\hat{X}})
\end{align*}
with
\begin{align*}
	\tau(\sigma_{\hat{X}}):=\sqrt{(1-e^{-2R})(1-e^{-2(R+R_c+P-\psi(\sigma_{\hat{X}}))})}.
\end{align*}
This is indeed the case when $R=0$. So it suffices to consider the case $R>0$.
For $\sigma_{\hat{X}}\in[\sigma(P),\sigma_X]$,
\begin{align*}
	\frac{\mathrm{d}\zeta(\sigma_{\hat{X}})}{\mathrm{d}\sigma_{\hat{X}}}&=2\sigma_{\hat{X}}-2\sigma_X\tau(\sigma_{\hat{X}})\\
	&\quad-\frac{2(1-e^{-2R})e^{-2(R+R_c+P-\psi(\sigma_{\hat{X}}))}(\sigma^2_X-\sigma^2_{\hat{X}})}{\sigma_X\tau(\sigma_{\hat{X}})}\\
	&\geq2\sigma_{\hat{X}}-2\sigma_X\sqrt{(1-e^{-2R})(1-e^{-2(R+R_c+P)})}\\
	&\quad-\frac{2\sqrt{1-e^{-2R}}e^{-2(R+R_c)}(\sigma^2_X-\sigma^2_{\hat{X}})}{\sigma_X\sqrt{1-e^{-2(R+R_c)}}}\\
	&=:f(\sigma_{\hat{X}}).
\end{align*}
Note that $f(\cdot)$ is a convex quadratic function with $f(0)<0$ and $f(\sigma_{X})>0$. So $f(\sigma_{\hat{X}})=0$ has a unique solution in $(0,\sigma_X)$, which can be shown to be $\sigma(R,R_c,P)$ given by (\ref{eq:sigma}). If $\sigma(P)\geq\sigma(R,R_c,P)$, then $\frac{\mathrm{d}\zeta(\sigma_{\hat{X}})}{\mathrm{d}\sigma_{\hat{X}}}\geq 0$ for $\sigma_{\hat{X}}\in[\sigma(P),\sigma_X]$, and consequently  $\zeta(\sigma_{\hat{X}})$ attaints its minimum over this interval at $\sigma_{\hat{X}}=\sigma(P)$. However, this is just a sufficient condition. A necessary condition for $\zeta(\sigma_{\hat{X}})$ to attaint its minimum  at $\sigma_{\hat{X}}=\sigma(P)$ is $\left.\frac{\mathrm{d}\zeta(\sigma_{\hat{X}})}{\mathrm{d}\sigma_{\hat{X}}}\right|_{\sigma_{\hat{X}}=\sigma(P)}\geq 0$, which can be expressed equivalently as $\sigma(P)\geq\varsigma(R,R_c)$ with $\varsigma(R,R_c)$ given by (\ref{eq:varsigma}).


Now we proceed to prove
(\ref{eq:sigma'}). Note that
\begin{align*} 
	f(\sigma_{\hat{X}})&\geq f'(\sigma_X)\\
	&:=2\sigma_{\hat{X}}-2\sigma_X\sqrt{1-e^{-2R}}\\
	&\quad-\frac{2\sqrt{1-e^{-2R}}e^{-2(R+R_c)}(\sigma^2_X-\sigma^2_{\hat{X}})}{\sigma_X\sqrt{1-e^{-2(R+R_c)}}}.
	\end{align*}
As $f'(\cdot)$ is a convex quadratic function with $f'(0)<0$ and $f'(\sigma_{X})>0$, the equation $f'(\sigma_{\hat{X}})=0$ has a unique solution in $(0,\sigma_X)$, which can be shown to be $\varsigma'(R,R_c)$ given by (\ref{eq:varsigma'}). Since $f(\varsigma'(R,R_c))\geq f'(\varsigma'(R,R_c))=0$, 
we must have $\varsigma'(R,R_c)\geq\sigma(R,R_c,P)$.

\section{Proof of Theorem \ref{thm:GaussianW2}}\label{app:GaussianW2}

		\begin{lemma}\label{lem:suffW2}
		For the case $\Delta(x,\hat{x})=(x-\hat{x})^2$ and $\phi(p_X,p_{\hat{X}})=W^2_2(p_X,p_{\hat{X}})$, we have
		\begin{align*}
			D(R,R_c,P|\mathcal{N}(\mu_X,\sigma^2_X))=&\inf\limits_{p_{\hat{X}}}D(R,R_c|\mathcal{N}(\mu_X,\sigma^2_X),p_{\hat{X}})\\
			\mbox{subject to}&\quad \mu_{\hat{X}}=\mu_X,\\
			&\quad\sigma_{\hat{X}}\leq\sigma_X,\\
			&\quad W^2_2(\mathcal{N}(\mu_X,\sigma^2_X),p_{\hat{X}})\leq P.
		\end{align*}
	\end{lemma}
	\begin{IEEEproof}
		The proof is similar to that of Lemma \ref{lem:suffKL}. It suffices to show
		\begin{align*}
			W^2_2(\mathcal{N}(\mu_X,\sigma^2_X),p_{\hat{X}'})\leq 	W^2_2(\mathcal{N}(\mu_X,\sigma^2_X),p_{\hat{X}}), 
		\end{align*}
		which is implied by (\ref{eq:variance}).
	\end{IEEEproof}
	
	In view of Lemma \ref{lem:suffW2}, it suffices to consider $p_{\hat{X}}$ with $\mu_{\hat{X}}=\mu_X$ and $\sigma_{\hat{X}}\leq\sigma_X$ for the purpose of computing $D(R,R_c,P|\mathcal{N}(\mu_X,\sigma^2_X))$.
	Further restricting $p_{\hat{X}}$ to be Gaussian and invoking Theorem \ref{thm:Gaussianoutputconstrained} 	
	yields the following upper bound on $D(R,R_c,P|\mathcal{N}(\mu_X,\sigma^2_X))$:
	\begin{align}
		&D(R,R_c,P|\mathcal{N}(\mu_X,\sigma^2_X))\nonumber\\
		&\leq\min\limits_{\sigma_{\hat{X}}\in[(\sigma_X-\sqrt{P})_+,\sigma_X]}\sigma^2_X+\sigma^2_{\hat{X}}-2\sigma_X\sigma_{\hat{X}}\xi(R,R_c).\label{eq:min}
	\end{align}
	Clearly,
	$\sigma^2_{\hat{X}}-2\sigma_X\sigma_{\hat{X}}\xi(R,R_c)$ is monotonically decreasing for $\sigma_{\hat{X}}\in[0,\sigma_X\xi(R,R_c)]$ and is monotonically increasing for $\sigma_{\hat{X}}\in[\sigma_X\xi(R,R_c),\infty)$. Therefore, the minimum  in (\ref{eq:min}) is attained at 
	$\sigma_{\hat{X}}=\sigma_X\xi(R,R_c)$
	when $\sigma_X-\sqrt{P}\leq\sigma_X\xi(R,R_c)$, and is attained at
	$\sigma_{\hat{X}}=\sigma(P)$
	when $\sigma_X-\sqrt{P}>\sigma_X\xi(R,R_c)$. This proves (\ref{eq:W2upper}).
	
	To prove (\ref{eq:W2lower}), we note that different from the case with Kullback-Leibler divergence, except when $P=0$, the constraint $W^2_2(\mathcal{N}(\mu_X,\sigma^2_{X}),p_{\hat{X}})\leq P$ does not imply any non-trivial lower bound on $h(\hat{X})$, and consequently a different approach is needed to bound $D(R+R_c|p_{\hat{X}})$.
	
	\begin{lemma}\label{lem:GaussianW2}
		For any $p_{\hat{X}}$ with $W^2_2(\mathcal{N}(\mu_X,\sigma^2_{X}),p_{\hat{X}})\leq P$, we have
		\begin{align*}
			D(R+R_c|p_{\hat{X}})\geq	(\sigma_Xe^{-(R+R_c)}-\sqrt{P-(\mu_X-\mu_{\hat{X}})^2})^2_+.
		\end{align*}
	\end{lemma}
	\begin{IEEEproof}
		Let $\hat{Y}$ be jointly distributed with $\hat{X}$ such that $I(\hat{X};\hat{Y})\leq R+R_c$. Construct a Markov chain $\tilde{X}\leftrightarrow\hat{X}\leftrightarrow\hat{Y}$, where $\tilde{X}\sim\mathcal{N}(\mu_X,\sigma^2_X)$ and $\mathbb{E}[(\tilde{X}-\hat{X})^2]=W^2_2(\mathcal{N}(\mu_X,\sigma^2_X),p_{\hat{X}})$. Note that
		\begin{align}
			&\mathbb{E}[((\tilde{X}-\mu_X)-(\hat{Y}-\mu_{\hat{Y}}))^2]\nonumber\\
			&=\mathbb{E}[((\tilde{X}-\mu_X)-(\hat{X}-\mu_{\hat{X}}))^2]\nonumber\\
			&\quad+\mathbb{E}[((\hat{X}-\mu_{\hat{X}})-(\hat{Y}-\mu_{\hat{Y}}))^2]\nonumber\\
			&\quad+2\mathbb{E}[((\tilde{X}-\mu_X)-(\hat{X}-\mu_{\hat{X}}))((\hat{X}-\mu_{\hat{X}})-(\hat{Y}-\mu_{\hat{Y}}))]\nonumber\\
			&\leq(\sqrt{\mathbb{E}[((\tilde{X}-\mu_X)-(\hat{X}-\mu_{\hat{X}}))^2]}\nonumber\\
			&\quad+\sqrt{\mathbb{E}[((\hat{X}-\mu_{\hat{X}})-(\hat{Y}-\mu_{\hat{Y}}))^2]})^2\nonumber\\
			&\leq(\sqrt{\mathbb{E}[((\tilde{X}-\mu_X)-(\hat{X}-\mu_{\hat{X}}))^2]}+\sqrt{\mathbb{E}[(\hat{X}-\hat{Y})^2]})^2\nonumber\\
			&=(\sqrt{W^2_2(\mathcal{N}(\mu_X,\sigma^2_X),p_{\hat{X}})-(\mu_X-\mu_{\hat{X}})^2}\nonumber\\
			&\quad+\sqrt{\mathbb{E}[(\hat{X}-\hat{Y})^2]})^2\nonumber\\
			&\leq(\sqrt{P-(\mu_X-\mu_{\hat{X}})^2}+\sqrt{\mathbb{E}[(\hat{X}-\hat{Y})^2]})^2.\label{eq:hand1}
		\end{align}
		On the other hand, we have
		\begin{align}
			\mathbb{E}[((\tilde{X}-\mu_X)-(\hat{Y}-\mu_{\hat{Y}}))^2]&\stackrel{(a)}{\geq}\sigma^2_Xe^{-2I(\tilde{X}-\mu_X;\hat{Y}-\mu_{\hat{Y}})}\nonumber\\
			&=\sigma^2_Xe^{-2I(\tilde{X};\hat{Y})}\nonumber\\
			&\stackrel{(b)}{\geq}\sigma^2_Xe^{-2I(\hat{X};\hat{Y})}\nonumber\\
			&\geq\sigma^2_Xe^{-2(R+R_c)},\label{eq:hand2}
		\end{align}
		where ($a$) and ($b$) are due to the Shannon lower bound \cite[Equation (13.159)]{CT91} and the data processing inequality \cite[Theorem 2.8.1]{CT91}, respectively. Combining (\ref{eq:hand1}) and (\ref{eq:hand2}) yields
		\begin{align*}
			\sqrt{\mathbb{E}[(\hat{X}-\hat{Y})^2]}\geq\sigma_Xe^{-(R+R_c)}-\sqrt{P-(\mu_X-\mu_{\hat{X}})^2},
		\end{align*}
		which, together with the fact $\mathbb{E}[(\hat{X}-\hat{Y})^2]\geq 0$, completes the proof of Lemma \ref{lem:GaussianW2}.
	\end{IEEEproof}

	Now we are in a position to prove (\ref{eq:W2lower}). Note that Proposition \ref{prop:W2}, together with the constraints $\mu_{\hat{X}}=\mu_X$, $\sigma_{\hat{X}}\leq\sigma_X$, and $W^2_2(\mathcal{N}(\mu_X,\sigma^2_X),p_{\hat{X}})\leq P$, implies
	\begin{align}
		\sigma_{\hat{X}}\in[(\sigma_X-\sqrt{P})_+,\sigma_X].\label{eq:domain}
	\end{align}
	According to \cite[Theorem 13.3.2]{CT91},
	\begin{align}
		D(R|\mathcal{N}(\mu_X,\sigma^2_X))=\sigma^2_Xe^{-2R}.\label{eq:term1}
	\end{align}
	Moreover, it follows by Lemma \ref{lem:GaussianW2} with $\mu_{\hat{X}}=\mu_X$ that
	\begin{align}
		D(R+R_c|p_{\hat{X}})\geq	(\sigma_Xe^{-(R+R_c)}-\sqrt{P})^2_+.\label{eq:term2}
	\end{align}
	In view of (\ref{eq:domain})--(\ref{eq:term2}),  invoking Corollary \ref{cor:generallowerbound} and (\ref{eq:remark}) yields the following lower bound:
	\begin{align*}
		D(R,R_c,P|\mathcal{N}(\mu_X,\sigma^2_X))\geq\min\limits_{\sigma_{\hat{X}}\in[(\sigma_X-\sqrt{P})_+,\sigma_X]}g(\sigma_{\hat{X}}),
	\end{align*}
	where 
	\begin{align*}
		g(\sigma_{\hat{X}})&:=\sigma^2_X+\sigma^2_{\hat{X}}\\
		&-2\sigma_X\sqrt{(1-e^{-2R})(\sigma^2_{\hat{X}}-(\sigma_Xe^{-(R+R_c)}-\sqrt{P})^2_+)}.
	\end{align*}
	
	It is clear that 
	\begin{align*}
		\min\limits_{\sigma_{\hat{X}}\in[(\sigma_X-\sqrt{P})_+,\sigma_X]}g(\sigma_{\hat{X}})=\sigma^2_X+(\sigma_X-\sqrt{P})^2_+
	\end{align*}
	when $R=0$.
	Henceforth we shall assume $R>0$. 
	First consider the case $\sqrt{P}\geq\sigma_Xe^{-(R+R_c)}$. In this case,
	\begin{align*}
		g(\sigma_{\hat{X}})=\sigma^2_X+\sigma^2_{\hat{X}}-2\sigma_X\sigma_{\hat{X}}\sqrt{1-e^{-2R}},
	\end{align*}
	which is monotonically decreasing for $\sigma_{\hat{X}}\in[0,\sigma_X\sqrt{1-e^{-2R}}]$ and is monotonically increasing for $\sigma_{\hat{X}}\in[\sigma_X\sqrt{1-e^{-2R}},\infty)$. Therefore, the minimum value of $g(\sigma_{\hat{X}})$ for $\sigma_{\hat{X}}\in[(\sigma_X-\sqrt{P})_+,\sigma_X]$ is attained at $\sigma_{\hat{X}}=\sigma_X\sqrt{1-e^{-2R}}$ if $\sqrt{P}\geq\sigma_X(1-\sqrt{1-e^{-2R}})$ and is attained at $\sigma_{\hat{X}}=\sigma_X-\sqrt{P}$ if $\sqrt{P}<\sigma_X(1-\sqrt{1-e^{-2R}})$, which, together with the assumption $\sqrt{P}\geq\sigma_Xe^{-(R+R_c)}$, gives
	\begin{align*}
		&\min\limits_{\sigma_{\hat{X}}\in[(\sigma_X-\sqrt{P})_+,\sigma_X]}g(\sigma_{\hat{X}})\\
		&=\begin{cases}
			\sigma^2_Xe^{-2R}&\hspace{-2.2in}\mbox{if }\frac{\sqrt{P}}{\sigma_X}\geq(1-\sqrt{1-e^{-2R}})\vee e^{-(R+R_c)},\\
			\sigma^2_X+(\sigma_X-\sqrt{P})^2-2\sigma_X(\sigma_X-\sqrt{P})\sqrt{1-e^{-2R}}\\
			&\hspace{-2.2in}\mbox{if }\frac{\sqrt{P}}{\sigma_X}\in[e^{-(R+R_c)},1-\sqrt{1-e^{-2R}}).
		\end{cases}
	\end{align*}
	Next consider the case $\sqrt{P}<\sigma_Xe^{-(R+R_c)}$. In this case, 
	\begin{align*}
		\frac{\mathrm{d}g(\sigma_{\hat{X}})}{\mathrm{d}\sigma_{\hat{X}}}=\sigma_{\hat{X}}\rho(\sigma_{\hat{X}}),
	\end{align*}
	where
	\begin{align*}
		\rho(\sigma_{\hat{X}}):=2-\frac{2\sigma_X(1-e^{-2R})}{\sqrt{(1-e^{-2R})(\sigma^2_{\hat{X}}-(\sigma_Xe^{-(R+R_c)}-\sqrt{P})^2)}}.
	\end{align*}
	It can be verified that $\rho(\sigma_{\hat{X}})\geq 0$ if and only if
	$\sigma_{\hat{X}}\geq\sqrt{\sigma^2_X(1-e^{-2R})+(\sigma_Xe^{-(R+R_c)}-\sqrt{P})^2}$. Moreover, $\sigma_X-\sqrt{P}>\sqrt{\sigma^2_X(1-e^{-2R})+(\sigma_Xe^{-(R+R_c)}-\sqrt{P})^2}$ if and only if $\frac{\sqrt{P}}{\sigma_X}<\nu(R,R_c)$,
	where $\nu(R,R_c)$ is defined in (\ref{eq:nu}). 	
	Therefore, under the assumption $\sqrt{P}<\sigma_Xe^{-(R+R_c)}$, the minimum value of $g(\sigma_{\hat{X}})$ for $\sigma_{\hat{X}}\in[(\sigma_X-\sqrt{P})_+,\sigma_X]$ is
	attained at $\sigma_{\hat{X}}=\sqrt{\sigma^2_X(1-e^{-2R})+(\sigma_Xe^{-(R+R_c)}-\sqrt{P})^2}$ if $\frac{\sqrt{P}}{\sigma_X}\geq\nu(R,R_c)$ and is attained at $\sigma_{\hat{X}}=\sigma_X-\sqrt{P}$ if $\frac{\sqrt{P}}{\sigma_X}<\nu(R,R_c)$, which gives
	\begin{align*}
		&\min\limits_{\sigma_{\hat{X}}\in[(\sigma_X-\sqrt{P})_+,\sigma_X]}g(\sigma_{\hat{X}})\\
		&=\begin{cases}
			\sigma^2_Xe^{-2R}+(\sigma_Xe^{-(R+R_c)}-\sqrt{P})^2\\
			&\hspace{-1.5in}\mbox{if }\frac{\sqrt{P}}{\sigma_X}\in[\nu(R,R_c),e^{-(R+R_c)}),\\
			\sigma^2_X+(\sigma_X-\sqrt{P})^2-2\sigma^2_X\sqrt{(1-e^{-2R})}\\
			\times\sqrt{(1-e^{-(R+R_c)})(1+e^{-(R+R_c)}-\frac{2\sqrt{P}}{\sigma_X})}\\
			&\hspace{-1.5in}\mbox{if }\frac{\sqrt{P}}{\sigma_X}<\nu(R,R_c)\wedge e^{-(R+R_c)}.
		\end{cases}
	\end{align*}



\ifCLASSOPTIONcaptionsoff
  \newpage
\fi

\end{document}